\documentclass[12pt]{article}
\usepackage{amsmath}
\usepackage{amstext,amssymb,amsfonts,
  graphicx,color,pstricks,pst-coil,pst-plot,psfrag,bbm} 

\advance\voffset by -1.5cm
\advance\hoffset by -1.25cm
\definecolor{darkgreen}{rgb}{0,0.6,0}
\def\thefootnote{\fnsymbol{footnote}}

\def\be{\begin{equation}}
\def\ee{\end{equation}}
\def\ba{\begin{eqnarray}}
\def\ea{\end{eqnarray}}

\newcommand{\N}{{\cal N}}

\newcommand{\CC}{\mathbb{C}}
\newcommand{\ZZ}{\mathbb{Z}}
\newcommand{\1}{\mathbbm{1}}

\newcommand{\ie}{{\it i.e.~}}

\newcommand{\nn}{{\nonumber}}

\newcommand{\RR}{{\mathbb R}}

\newcommand{\mk}{{\mathcal{K}^{(6)}}}
\newcommand{\ck}{C^{(6)}}

\usepackage{amsmath,amsfonts,epsfig,color,latexsym}
\usepackage{amssymb}
\usepackage[english, USenglish]{babel}
\usepackage{epsfig}
\usepackage{rotating}
\usepackage{slashed}

\renewcommand{\theequation}{\arabic{section}.\arabic{equation}}

\usepackage[unicode=true, ps2pdf, dvips, linktocpage, bookmarks=true]{hyperref}
\begin{document}


\thispagestyle{empty}
\renewcommand{\thefootnote}{\fnsymbol{footnote}}

{\hfill \parbox{4cm}{
}}

\bigskip\bigskip

\begin{center} \noindent \Large \bf
Heterotic $\mathbf{AdS_3/CFT_2}$ duality\\
with $\mathbf{(0, 4)}$ spacetime supersymmetry
\end{center}

\bigskip\bigskip\bigskip

\centerline{ \normalsize \bf Stefan Hohenegger, Christoph A.\ Keller 
and Ingo Kirsch$^a$ \footnote[1]{\noindent \tt
 email: stefanh@phys.ethz.ch, kellerc@phys.ethz.ch, kirsch@phys.ethz.ch}}

\bigskip

\centerline{\it ${}^a$ Institut f\"ur Theoretische Physik, ETH
  Z\"urich}
\centerline{\it 
CH-8093 Z\"urich, Switzerland}

\bigskip\bigskip

\bigskip\bigskip

\renewcommand{\thefootnote}{\arabic{footnote}}

\bigskip\bigskip

\centerline{\bf \small Abstract}
\medskip

{\small \noindent We discuss the AdS$_3$/CFT$_2$ duality of a
  heterotic three-charge model with $(0, 4)$ target space
  supersymmetry. The worldsheet theory for heterotic strings on the
  $AdS_3 \times S^3/\ZZ_N \times T^4$ near-horizon geometry was
  constructed by Kutasov, Larsen and Leigh in [hep-th/9812027]. We
  propose that the dual conformal field theory is given by a
  two-dimensional $(0, 4)$ sigma model arising on the Higgs branch of
  an orbifolded ADHM model. As a non-trivial consistency check of the
  correspondence, we find that the left- and right-moving central
  charges of the infrared conformal field theory agree with those
  predicted by the worldsheet model. Moreover, using the entropy
  function formalism, we show that to next-to-leading order the 
  central charge can also be obtained from an $\alpha'$-corrected
  supergravity theory. }


\newpage
\setcounter{tocdepth}{2}
\tableofcontents

\setcounter{equation}{0}
\section{Introduction}

Recently, several authors \cite{Giveon:2006pr, TalkStrominger,
  Dabholkar, Johnson2, Strominger, Kraus, Alishahiha2008} have studied
the possibility of an AdS$_3$/CFT$_2$ duality for the fundamental {\em
  heterotic} string.  Heterotic strings are dual to type~I D1-branes
whose low-energy effective field theory is expected to be conformally
invariant. The dual near-horizon geometry of the heterotic string
should therefore contain an $AdS_3$ factor. This was confirmed in
\cite{CDKL} (see also \cite{Castro:2008ne}) in which an $AdS_3 \times
S^2$ factor was found in a $\N=2$, $d=5$ $R^2$-corrected supergravity
solution corresponding to heterotic strings in five dimensions.

In general, heterotic string setups may contain additional charged
objects such as NS5-branes and Kaluza-Klein monopoles.  Such setups
generically have $(0, 4)$ target space supersymmetry.  Recently, it
has been found that in the absence of some or all of these additional
charges the target space supersymmetry is enhanced to $(0,8)$
\cite{TalkStrominger, Johnson2, Strominger, Kraus} (see also
\cite{Alishahiha2007}).  Such theories are expected to be very
different from those with only $(0, 4)$ supersymmetry. For one thing,
there are no linear superconformal algebras with more than four
supercurrents.  Indeed, it has been argued in \cite{Strominger, Kraus}
that the global supergroup of the boundary CFT is $Osp(4^*|4)$, whose
affine extension is given by a {\em nonlinear} $\N = 8$, $d=2$
superconformal algebra. For another, it is not clear if these theories
possess unitary representations.

In this paper we take a step back and address the construction of a
heterotic AdS/CFT duality with only $(0, 4)$ target space
supersymmetry.  For this we revisit a heterotic three-charge model
previously studied by Kutasov, Larsen and Leigh (KLL) in
\cite{KLL}. The setup consists of $p$ fundamental strings embedded in
the worldvolume of $N'$ NS5 branes and $N$ Kaluza-Klein (KK)
monopoles. In \cite{KLL} KLL work out the worldsheet theory for string
theory on the corresponding near-horizon geometry $AdS_3
\times S^3/\ZZ_N \times T^4$. The worldsheet CFT turns out to be
essentially the product of an $SL(2)$ WZW model and a ``twisted''
$SU(2)$ WZW model corresponding to the asymmetric orbifold
$S^3/\ZZ_N$. In contrast, not much is known about the dual conformal
field theory on the boundary of the $AdS_3$ space.

The first part of this paper is therefore devoted to the construction
of the dual two-dimensional boundary conformal field theory. We first
apply heterotic/type~I duality to map the three-charge configuration
to an intersection of $p$ D1-branes and $N'$ D5-branes plus $N$ KK
monopoles in type~I string theory. In the absence of any KK monopoles
the low-energy effective theory corresponds to Witten's ADHM sigma
model of Yang-Mills instantons \cite{Witten:1994tz}, as shown by
Douglas in \cite{Douglas}. To also include KK monopoles, which have a
$\CC^2/\ZZ_N$ near-core geometry, it is natural to construct a $\ZZ_N$
orbifold theory of the massive ADHM sigma model. (Refs.~\cite{Sugawara,
  Okuyama:2005gq} also use an orbifold construction to obtain the
boundary CFT dual to {\em type II} string theory on $AdS_3 \times
S^3/\ZZ_N \times T^4$.)
 
Our proposal is that the sought-after boundary conformal field theory
arises on the Higgs branch of the orbifolded ADHM model, which
corresponds to the bound state phase of the D-brane setup. We will
perform a consistency check for the proposal by the following line of
reasoning. Lambert has shown in \cite{Lambert2} that, even though the
ADHM model is classically not conformal, it is ultraviolet finite to
all orders in perturbation theory. There is no renormalisation group
flow, and anomalous conformal dimensions are absent \cite{Lambert2}.
The conformal Higgs branch theory can therefore be obtained by
integrating out the massive degrees of freedom in the ADHM model
\cite{Lambert}.  Moreover, the central charges of the Higgs branch
theory can be determined by counting the massless degrees of freedom
of the ultraviolet theory. In other words, they are given by the
dimension of the instanton moduli space of the ADHM model.  Repeating
these steps for the orbifold version of the ADHM model, we determine
the central charges of the low-energy theory of the three-charge model
and match them to those predicted by the worldsheet theory.

The second part of the paper is devoted to the construction of a
higher-derivative correction of the near-horizon supergravity solution
of the KLL setup.  In fact, for a dual setup a full solution of the
$\N=2$ off-shell completion of four-derivative supergravity in five
dimensions was constructed already in \cite{CDKL}. Here we will use
six-dimensional corrections to the heterotic string action
\cite{Metsaev:1987zx,Hull:1987pc} and employ the entropy function
formalism \cite{Sen:2005wa,Sen:2005iz} to find the corrected
near-horizon geometry. To first order, the latter correctly reproduces
the expected central charges of the boundary CFT via the
Brown-Henneaux formula \cite{BH}.

\section{Heterotic \texorpdfstring{$AdS_3/CFT_2$}{} duality}

In this section we review the supergravity solution of the heterotic
three-charge model of~\cite{KLL} and the corresponding worldsheet
model. Readers familiar with Ref.~\cite{KLL} may wish to proceed
directly to the discussion of the boundary conformal field theory in
section~\ref{bcftmodel}.

\subsection{Three-charge model for heterotic strings} \label{sec21}

We consider heterotic string theory compactified on $S^1\times T^4$
which we take along the directions $\{x^5\}$ and $\{x^6,x^7,x^8,x^9\}$
respectively. In particular, following \cite{KLL}, we study
the following brane setup:
\begin{itemize}
\item $p$ fundamental strings F1 infinitely stretched in the $x^1$ 
direction,
\item $N'$ NS5-branes wrapped around the $T^4$ and infinitely stretched 
along $x^1$,
\item $N$ KK monopoles wrapped around $T^4$ and extended in $x^1$.
\end{itemize}
We can depict this configuration schematically in the following table:\\
\begin{center}
\begin{tabular}{|c|c||cc|ccc|c|cccc|}
\hline
 &&0&1&2&3&4&5&6&7&8&9\\
\hline
 $p $  &F1  &$\bullet$&$\bullet$&&&&&&&& \\
 $N'$ &NS5 &$\bullet$&$\bullet$&&&&&$\bullet$&$\bullet$&$\bullet$&$\bullet$\\
 $N$ & KKM &$\bullet$&$\bullet$&&&&&$\bullet$&$\bullet$&$\bullet$&$\bullet$\\
\hline
\end{tabular}
\end{center}
$\ $\\[10pt] From a 5-dimensional spacetime point of view this
configuration looks like an infinitely stretched string in the $x^1$
direction, which preserves $(0, 4)$ supersymmetry, {\em i.e.}\ it is
non-supersymmetric in the left sector and contains four supercharges
in the right sector. Let us recall the classical solution as 
given in \cite{KLL}. The metric is given by
\begin{align}
ds^2=\,&F^{-1}(-dt^2+dx_1^2)+H_{5}\big[H_{K}^{-1}(dx_5+P_K(1-\cos\theta)
d\varphi)^2 \nonumber\\ 
&+H_K(dr^2+r^2(d\theta^2+\sin^2\theta
d\varphi^2))\big]+\sum_{i=6}^9dx_i^2,\label{classicalmetric}
\end{align}
with the following harmonic functions
\begin{align}
&H_5=1+\frac{P_5}{r}\,, 
&&H_K=1+\frac{P_K}{r}\,, 
&&F=1+\frac{Q}{r}\,.\label{harmfunct}
\end{align}
Here we use spherical coordinates $(r,\theta,\varphi)$ for the directions $(x^2,x^3,x^4)$. The
corresponding gauge fields and the dilaton read
\begin{align}
&B_{t1}=F\,,
&&B_{\varphi 5}=P_5(1-\cos\theta)\,,
&&e^{-2[\Phi_{10}(r)-\Phi_{10}(\infty)]}=\frac{F}{H_5} \,.
\end{align}
The quantities $P_5,P_K,Q$ are related to $N',N,p$ by
\begin{align}
 &P_5=\frac{\alpha'}{2R}N'\,, &&P_K=\frac{R}{2}N\,, 
&&  Q=\frac{{\alpha'}^3e^{2\Phi_{10}(\infty)}}{2RV}p\,,
\end{align}
where $R$ is the asymptotic radius of the $S^1$, $V$ the volume of the
torus and $\Phi_{10}(\infty)$ the asymptotic value of the dilaton.\\

\medskip

In the near-horizon limit $r\to 0$, the metric (\ref{classicalmetric})
reduces to
\begin{align}
ds^2=\,&\frac{{r'}^2}{4P_5P_K}(-dt^2+dx_1^2)
+\frac{P_5}{P_K}(dx_5+P_K(1-\cos\theta)d\varphi)^2\nonumber\\
&+P_5P_K\left(4d{r'}^2+(d\theta^2+\sin^2\theta d\varphi^2)\right)
+\sum_{i=6}^9 dx_i^2 \,, \label{AdSmetric}
\end{align}
where we have defined $r'$ by
\begin{align}
r=\frac{4P_5P_K{r'}^2}{Q} \,.
\end{align}
In \cite{KLL} this metric was interpreted as describing the space
\begin{align}
AdS_3\times S^3/\mathbb{Z}_N\times T^4 \,, 
\end{align}
with AdS radius and six-dimensional string coupling
\begin{align} \label{AdSrad}
R^{2}_{AdS, \, {\rm uncorr}} = \alpha' N N' \,,\qquad 
g^2_6 =e^{2\Phi_6^{\text{hor}}}= \frac{N'}{p} \,.
\end{align}
Obviously, string theory on this background is weakly-coupled for ${N'} \ll p$. Note that so far we have only discussed an {\em uncorrected}
supergravity solution, {\em i.e.}\ a solution to an action at the two
derivative level.  In section~\ref{EntropyFunction} we will address
the question of how to modify (\ref{AdSmetric}) in the presence of
higher derivative interactions.

\subsection{Lift to M-theory} \label{sec22}

In order to understand why the supergravity solution (\ref{AdSmetric})
is expected to receive $\alpha'$~corrections, we now determine the
central charges of the boundary CFT. We begin by mapping the heterotic
setup to M-theory compactified on $CY_3=K3 \times T^2$. For this, we
first dualize to type IIA theory, from where (after additional S and T
dualities) we may lift to M-theory --- see appendix
\ref{dualitiesMtheory} for details. We obtain the following
setup of M5 branes:\\
\begin{center}
\begin{tabular}{|c|c||ccc|cc|cccccc|}\hline
& & 0 & 1 & 2 & 3 & 4 & 5 & 6 & 7 & 8 & 9 & 10\\\hline
$p$ & M5 & $\bullet$ & $\bullet$ & & & & & $\bullet$ 
& $\bullet$ & $\bullet$ & $\bullet$ & \\
$N'$ & M5 & $\bullet$ & $\bullet$ & & & & $\bullet$ & $\bullet$ 
& $\bullet$ & & & $\bullet$\\
$N$ & M5 & $\bullet$ & $\bullet$ & & & & $\bullet$ & 
& & $\bullet$ & $\bullet$ & $\bullet$\\\hline
\end{tabular} 
\end{center}
$\ $\\
Our convention will be that the internal $T^2$ is spanned
by the directions $\{x^5,x^{10}\}$ while the $K3$ resides in
$\{x^6,x^7,x^8,x^9\}$.\\

\noindent
A general method for determining the central charges of the low-energy
effective theory on M5-branes wrapping a 4-cycle in a Calabi-Yau
three-fold $CY_3$ is given in \cite{Maldacena}. The low-energy
effective field theory is given by a two-dimensional (heterotic) sigma
model with the M5-brane moduli space as target space. The left- and right-moving central
charges $c_{L,R}$ of this sigma model are given by
\begin{align} 
 c_L &= 6 D +  c_2 \cdot p \,,\qquad 
 c_R = 6 D + \frac{1}{2} c_2 \cdot p \,,\label{cLR}  \nonumber\\
 D&=\frac{1}{6} c_{IJK} p^I p^J p^K \,,
\end{align}
where $c_{IJK}$ are the intersection numbers of $CY_3$, and $p^I$ is
the (magnetic) charge of the M5-brane wrapping the $I$th 4-cycle
\cite{Maldacena}. The product $c_2 \cdot p$ contains the second Chern
class of $CY_3$.\footnote{For an exact definition of the product $c_2
  \cdot p$ see \cite{Maldacena}.} \\

\noindent
Let us apply these formulae to the present case\footnote{In contrast
  to what is assumed in \cite{Maldacena} for the four-cycle inside the
$CY_3$, $K3$ is not a very ample divisor in $K3 \times
T^2$. Nevertheless, we may still use (\ref{cLR}), since $b_1(K3)=0$, 
even though $b_1(K3 \times T^2) \neq 0$.} and identify
\begin{align}
&p^1=p\,, &&p^2=N\,, &&p^3=N'\,.
\end{align}
Denoting the single modulus of the $T^2$ by~$p^1$, the only
non-vanishing intersection numbers are $c_{1ij}=c_{ij}$, where $c_{ij}$
is the intersection matrix for $K3$. For $p$ M5-branes wrapping $K3$,
$c_2 \cdot p= c_2(K3) p =24 p$ \cite{CDKL}, and (\ref{cLR}) provides
the central charges 
\begin{align} 
c_L &= 6 N N'p + 24 p \,, \nonumber\\
c_R &= 6 N N'p + 12 p \,.  \label{cc1} 
\end{align}
Since $D \neq 0$, this three-charge model preserves only $(0,4)$
supersymmetry~\cite{Maldacena}. For $N=N'=0$, we have $D=0$ and
$(c_L,c_R)=(24 p, 12p)$.  These are the central charges of the $(0,
8)$ low-energy effective field theory describing a stack of
$p$~heterotic strings. \\

\noindent
Let us compare the central charges $c_{L, R}$ with that obtained from
the supergravity solution by applying the Brown-Henneaux formula
\cite{BH}, 
\begin{align}
c=\frac{3R_{AdS}}{2G_N^{(3)}} \,, \label{cc}
\end{align}
where $G_N^{(3)}$ is Newton's constant in three dimensions.
Substituting the AdS radius (\ref{AdSrad}) of the uncorrected
supergravity solution into (\ref{cc}), we get
\begin{align}
c=6NN'p \,, \label{ccKLL}
\end{align}
as was already found in \cite{KLL}.  We notice that (\ref{ccKLL}) agrees
with (\ref{cc1}) only to leading order in the charges. The reason for
the absence of the subleading term in (\ref{ccKLL}) is the fact that
it is computed from an uncorrected supergravity solution.  Taking into
account higher derivative terms in the action as well, one recovers
the full expression (\ref{cc1}), as was recently shown for a dual
setup \cite{CDKL}. We will reproduce this result with somewhat different
methods in section~\ref{EntropyFunction}.

\subsection{\texorpdfstring{$\N=(0,2)$}{} worldsheet theory} \label{secws}

We now discuss heterotic string theory on the $AdS_3 \times
S^3/\ZZ_{N} \times T^4$ near-horizon geometry of the F1-NS5-KKM
three-charge model introduced in section~\ref{sec21}.  The
corresponding worldsheet theory has been constructed in \cite{KLL},
and we will only review some of its features relevant for the
construction of the boundary conformal field theory.

The worldsheet theory is expected to be the product of a heterotic
$SL(2)$ WZW model, a conformal field theory on $S^3/\ZZ_{N}$ and a
free $U(1)^4$ CFT on the four-torus~$T^4$.  As a heterotic model, the
product theory is bosonic in the left-moving sector and supersymmetric
in the right-moving sector.  The heterotic $SL(2)$ WZW model therefore
has a bosonic affine $SL(2)$ algebra of level $k_b$ in the left-moving
sector and a supersymmetric one of level $k_s=k_b-2$ in the
right-moving sector.  Accordingly, the right-moving sector is
generated by three bosonic and three fermionic currents, $\bar J^A$
and $\bar \psi^A$ ($A=1,2,3$), while the left-moving sector contains
only $J^A$. Similarly, the right-moving CFT on
$T^4$ is constructed from four bosonic fields $\bar Y^i$ and 
four fermions $\bar \lambda_i$ ($i=1,2,3,4$).  The left-moving sector
contains only the bosonic currents $Y^i$.

In the unorbifolded case, the $S^3$ factor of the geometry would
be described by an $SU(2)$ WZW model with levels $k'_b$ and
$k'_s=k'_b+2$ in the left- and right-moving sector, respectively.  The
right-moving sector of the $SU(2)$ model contains three bosonic
currents, $\bar K^a$, and three fermions $\bar \chi^a$ $(a=1,2,3)$.
The left-moving sector has the same bosonic currents $K^a$
$(a=1,2,3)$, but again no fermions.

Let us now implement the $\ZZ_{N}$
orbifold.  We start from the
$SU(2)$ WZW model in which we parameterise the $SU(2)$ group manifold
in terms of the Euler angles
\begin{align}
0 \leq \theta \leq \pi\,,\qquad 0 \leq \phi \leq 2\pi\,, \qquad
0 \leq \xi \leq 4\pi \,, \label{xi}
\end{align}
where $\xi$ parameterises the fibre, and $\theta, \phi$ are the base
coordinates.  As in \cite{KLL}, we consider an $SU(2)$ model at level
\begin{align}
k'_b=N N' \label{k2b}
\end{align}
and identify
\begin{align}
\xi \sim \xi +\frac{4\pi}{N} \,.
\end{align}
The orbifold acts asymmetrically in the near-horizon geometry. We
therefore turn the $SU(2)$ WZW model into a coset model of the
type
\begin{align}
  \frac{SU(2)_L \times SU(2)_R}{(\ZZ_{N})_L} \,,
\end{align}
where the orbifold is embedded in $SU(2)_L$: $\ZZ_{N}$ 
acts on the currents as
\begin{align}
K^\pm \rightarrow e^{\pm \frac{4\pi i}{N}}K^\pm \,,&\qquad K^3 
\rightarrow K^3 \,, \nonumber\\
\bar K^{\pm, 3} \rightarrow \bar K^{\pm, 3}  \,,&\qquad \bar \chi^{\pm, 3} \
\rightarrow \bar \chi^{\pm, 3}  \,.
\end{align}
For $N>2$, the effect of the asymmetric orbifold is to break the $SU(2)$
of the left-moving sector down to~$U(1)$, whose current $K^3$ is
invariant under the orbifold action.\footnote{In the related $S^2$
  theory of \cite{Giddings} the orbifold is embedded in the
  supersymmetric (right) sector, and $(0, 2)$ worldsheet supersymmetry
  relates $N$ and $N'$. In the present case $N$ and $N'$ are
  independent since the orbifold is embedded in the
  non-supersymmetric (left) sector.}

\medskip

The consistency of the theory requires that the
worldsheet central charges are $(c^{\rm{ws}}_L, c^{\rm{ws}}_R)=(26,
15)$.  The central 
charges in the right-moving sector are
\begin{align}
c^{\rm{ws}}_R (AdS_3) = \frac{3}2 + \frac{3k_b}{k_b-2} \,,\qquad
c^{\rm{ws}}_R (S^3/\ZZ_{N}) = \frac{3}{2} + \frac{3k_b'}{k'_b+2} \,,\qquad
c^{\rm{ws}}_R (T^4) = 6 \,,
\end{align}
which adds up to $c^{\rm{ws}}_R=15$ provided that
\begin{align}
k_b=k'_b+4 \,. \label{kb}
\end{align}
Similarly, for the left-moving sector we have
\begin{align}
c^{\rm{ws}}_L (AdS_3) = \frac{3k_b}{k_b-2} \,,\qquad
c^{\rm{ws}}_L (S^3/\ZZ_{N}) = \frac{3k_b'}{k'_b+2} \,,\qquad
c^{\rm{ws}}_L (T^4) = 4 \,,
\end{align}
which adds up to ten. Heterotic string theory also contains 32
left-moving current algebra fermions, {\em i.e.}\ 16 for each $E_8$.
We thus get $c^{\rm{ws}}_L = 10+16=26$, as required.

\medskip The worldsheet theory also provides some information on the
boundary conformal field theory. As shown in \cite{Seiberg}, the left-
and right-moving (super)Virasoro algebras of the boundary CFT can be
constructed from the worldsheet affine $SL(2)$ Lie algebra. 
Their central charges are
\begin{align}
(c_{L}, c_R)  = (6 k_b p, 6 k_{s} p) \,, \label{cc2}
\end{align} 
where, as before, $k_b$ and $k_s=k_b-2$ are the levels of left- and
right-moving $SL(2)$ algebras, and $p$ is the number of heterotic
strings. Substituting (\ref{k2b}) and (\ref{kb}) in (\ref{cc2}), we
find the central charges
\begin{align}\label{cc3}
 (c_L, c_R)&=(24p + 6 N N' p, 12 p + 6 N N' p)  \,
\end{align}
which agree with (\ref{cc1}) and satisfy the constraint \ $c_L-c_R=12
p$ as also found in \cite{KLL, Kraus}. 

\medskip

Let us finally consider the amount of worldsheet and target space
supersymmetry. From the geometry we expect that the worldsheet model
preserves a $(0, 4)$ target space supersymmetry.  Since $T^4$ is
K\"ahler, the heterotic worldsheet CFT on $T^4$ has $(0, 2)$
supersymmetry. The K\"ahler structure also ensures that the $(0, 2)$
worldsheet supersymmetry leads to $(0, 4)$ spacetime supersymmetry.
The heterotic $SL(2)$ model and the ``twisted'' $SU(2)$ model
separately preserve only $(0, 1)$ supersymmetry. Only the product of
both models has a chance to have $(0, 2)$ worldsheet supersymmetry. In
order to enhance $\N=1$ to $\N=2$ supersymmetry in the right sector,
one must find a $U(1)_R$ current $J_{\N=2}$, which is part of the
$\N=2$ algebra. The existence of such a current is guaranteed by the
fact that the orbifold is embedded in $SU(2)_L$ such that the
right sector remains unaffected by it. The $\N=2$ $U(1)_R$ current
therefore has the same structure as in the (unorbifolded) type II
case, see \cite{Seiberg}.

\setcounter{equation}{0}
\section{Two-dimensional boundary sigma model}\label{bcftmodel}

\subsection{General remarks}\label{secgeneralrem}
In this section we discuss the two-dimensional $(0, 4)$ conformal
field theory living on the boundary of the $AdS_3$ space.  Our
starting point is the heterotic brane setup introduced in the previous
section. We first T-dualize in $x^5$ to go from $E_8 \times E_8$ to
$SO(32)$ heterotic string theory and then use heterotic/type~I duality
in order to obtain the following type~I brane configuration:

\begin{center}
\begin{tabular}{|c|c||cc|cccc|cccc|}
\hline
 &&0&1&2&3&4&5&6&7&8&9\\
\hline
 $p$  &D1 &$\bullet$&$\bullet$&&&&&&&&  \\
 $N'$ &D5 &$\bullet$&$\bullet$&&&&&$\bullet$&$\bullet$&$\bullet$&$\bullet$\\
 $N$ & KKM &$\bullet$&$\bullet$&&&&&$\bullet$&$\bullet$&$\bullet$&$\bullet$\\
 32 & D9 & $\bullet$&$\bullet$&$\bullet$&$\bullet$&$\bullet$&
           $\bullet$&$\bullet$&$\bullet$&$\bullet$&$\bullet$\\
\hline
\end{tabular}
\end{center}
Let us consider the type~I setup in detail. Since the heterotic/type~I
duality involves a strong-coupling transition, the heterotic F1 and
NS5-branes naturally map to D1 and D5-branes.  Moreover, since we are
dealing with a type~I string theory we are also required to introduce
32 D9-branes and perform an orientifold projection. In order to
understand the contribution of the KK monopoles, we recall that the
approximation of the near-core region of $N$ KK monopoles is a
$\CC^2/\ZZ_N$ orbifold. This instructs us to study a $\ZZ_N$ orbifold
in the directions $x^{2,3,4,5}$ of the D1-D5-D9-brane theory.\\

\noindent
In order to set up our notation we remark that the D1-D5-D9 brane
configuration breaks ten-dimensional Lorentz symmetry to
$SO(1,1)\times SO(4)_E \times SO(4)_I$, where $SO(4)_E$ and $SO(4)_I$
rotate $x^{2,3,4,5}$ and $x^{6,7,8,9}$, respectively. We will
use the standard decomposition 
\begin{equation*}
SO(4)_E \times SO(4)_I \simeq 
  SU(2)_{A} \times SU(2)_{Y} \times SU(2)_{A'} \times SU(2)_{\tilde A'}\,
\end{equation*}
to label the appearing representations in terms of doublet
representations with ($A', \tilde A'$, $A, Y = \pm$).  The orbifold 
is embedded in $SU(2)_Y$.\\

\noindent
We will start out our construction by reviewing the low-energy
effective theory of the type~I D1-D5-D9 intersection.  In the absence
of any KK monopoles this theory was shown in~\cite{Douglas} (for
$p=1$) to be equivalent to Witten's ADHM model of Yang-Mills
instantons. In section~\ref{secADHM} we will review the model for
$p>1$ as constructed in \cite{Lowe}. In section~\ref{secorbifold} we will
include the effect of the KK monopoles by orbifolding the ADHM
model. Subsequently, in section~\ref{secinst} we discuss its instanton
moduli space and determine the central charges of the Higgs branch
theory.

\subsection{Spectrum of D1-D5-D9 and the ADHM model}
\label{secADHM}  

Let us briefly recall some basic facts. Spacetime fermions arise in
the Ramond sector, and spacetime bosons in the Neveu-Schwarz sector.
If the boundary conditions on both ends of the string are the same,
then the worldsheet fermions of the R sector have integral modes, and
those in the NS sector half-integers.  If the boundary conditions are
different, the additional signs introduced exchange the moddings, which
also changes the ground state energy of the sector. In particular, the NS
ground state energy in the case of $N_{DN}$ mixed boundary conditions
is given by \be -\frac{1}{2}+\frac{N_{DN}}{8}\ , \label{Egs} \ee
whereas the ground state energy in the R sector is always
zero.\\

\noindent
Let us now discuss the strings stretching between the various types of
branes.

\subsubsection*{1-1 strings}
In the NS-sector, the massless modes form a ten-dimensional
vector $A^\mu_{ab}$, the Chan-Paton indices running over $a,b=1,\ldots,
p$.  Considered as an object on the D1, it splits into a 2d vector
$A^\mu_{ab}$ and 8 scalars $b^i_{ab}$. The orientifold projection
$\Omega$ maps $A^\mu_{ab}\mapsto -A^\mu_{ba}$. We are thus left with the
gauge bosons $A^\mu_{[ab]}$ in the adjoint of the gauge group
$SO(p)$. On the other 
hand, the vertex operator of $b$ picks up no sign under $\Omega$, as
it contains no derivative along the boundary. This leaves 8 bosons
$b^i_{(ab)}$ in the symmetric representation of $SO(p)$ which we group
in a pair of 4 bosons, $b^{AY}_{(ab)}$ and $b^{A'\tilde
A'}_{(ab)}$.

In the R-sector, the GSO projection restricts to modes which are 
invariant under $\bar \Gamma := \Gamma^0 \ldots \Gamma^9$, where
$\Gamma^\mu$ denotes the fermionic zero modes. To obtain the action of
$\Omega$, note that the fermionic modes $\psi^2, \ldots, \psi^9$
reflect from the boundary with an extra minus sign, so that they pick
up an additional minus sign under exchange of right and left
movers. $\Omega$ thus acts on massless fermions as $\Omega = -\Gamma^2
\Gamma^3 \ldots \Gamma^9$. The massless spinors thus must satisfy the
two conditions
\be
\psi_{ab} = \bar\Gamma \psi_{ab} = -\Gamma^2\ldots\Gamma^9 \psi_{ba}\
 . \label{2cond}
\ee
The first condition simply states that $\psi$ is in the {\bf 16} of
$SO(1,9)$. To obtain the worldsheet behaviour of $\psi$, we need to
decompose {\bf 16} into representations of $SO(1,1)\times SO(8)$,
which gives $\mathbf{16}=\mathbf{8}'_+ \oplus \mathbf{8}''_-$, where
{$\mathbf{8}'$, $\mathbf{8}''$} are the two spinor representations of
$SO(8)$ and $\pm$ denotes the chirality with respect to $SO(1,1)$. The
second condition in (\ref{2cond}) then states that $\psi_{(ab)}$
transforms as $\mathbf{ 8}''_-$, and $\psi_{[ab]}$ as $\mathbf{
8}'_+$. The $\psi_{(ab)}$ are the right-moving superpartners of the
$b_{(ab)}$. Due to the D5-branes each ${\mathbf{8}}$ decomposes into a
pair of $\mathbf{4}$'s of the $SO(4)$'s. Following \cite{Douglas},
these will be denoted by $\psi^{A'Y}_{-\,(ab)}$,$\psi^{A\tilde
A'}_{-\,(ab)}$ and $\psi^{A'A}_{+\,[ab]}$, $\psi^{Y\tilde
A'}_{+\,[ab]}$. The left-moving fermions $\psi_{+\,[ab]}$ are
antisymmetric and therefore do not appear in the case of a single
D1-brane.

\subsubsection*{1-5 strings}
The analysis of this sector has been performed in
\cite{Douglas:1995bn}.  Since $N_{DN}=4$, the ground state energy is
also zero in the NS-sector, so that there appear both bosons and
fermions. In total, we obtain 
bosons $\phi_a^{A'm}$ in the $(p,2N',1)$ of $SO(p)\times Sp(2N')\times
SO(32)$, and their right- and left-moving fermionic superpartners
$\chi_{-\,a}^{Am}$ and $\chi_{+\,a}^{Ym}$. The index $m$ runs over
$m=1,...,2N'$.

\subsubsection*{1-9 strings}
Since $N_{DN}=8$, the ground state energy of the NS-sector is strictly
positive, so that there are no bosons.  In the R-sector there are two
massless modes $\Gamma^0, \Gamma^1$.  The GSO projection eliminates
one of them, leaving only the left moving mode. We thus obtain $32 p$
left-moving fermions $\lambda_{+a}^M$, where $M=1,...,32$ is the
Chan-Paton index of $SO(32)$.

\subsubsection*{5-5 strings, 5-9 strings}
The analysis of the remaining sectors has been performed in
\cite{Douglas}.  Since their field content is not very important in
what follows, we only cite the results. The 5-brane fields form a
$Sp(2N')$ gauge theory, a hypermultiplet in the antisymmetric
representation with scalar component $X^{AY}_{[mn]}$, and
``half-hypermultiplets'' in $(1,2N',32)$ with scalar component
$h_M^{Am}$.\\

\noindent We summarise the results by listing the relevant fields in the following table (see also \cite{Lowe}):\\
\begin{table}[ht]
\begin{center}
\begin{tabular}{|c|c|c|c|}
\hline
strings & bosons & fermions & $SO(p)$ rep.\\ 
\hline
1-1& $A^\mu_{[ab]}$ & \parbox{2.3cm}{\vspace{0.1cm}$\psi^{A'A}_{+\,[ab]}$, $\psi^{Y\tilde A'}_{+\,[ab]}$\vspace{0.1cm}}& 
adj.$=$anti-sym.\\
&$b^{AY}_{(ab)}$  & $\psi^{A'Y}_{-\,(ab)}$ & sym.\\ 
&$b^{A'\tilde A'}_{(ab)}$  & \parbox{1cm}{\vspace{0.1cm}$\psi^{A\tilde A'}_{-\,(ab)}$\vspace{0.1cm}}& sym. \\
\hline
1-5&$\phi^{A'm}_a$ &\parbox{0.8cm}{\vspace{0.1cm}$\chi^{Am}_{-\,a}$\vspace{0.1cm}} & fund. \\ 
&&\parbox{0.8cm}{\vspace{0.1cm}$\chi^{Ym}_{+\,a}$\vspace{0.1cm}} & fund. \\
\hline
1-9 &&\parbox{0.8cm}{\vspace{0.1cm}$\lambda^M_{+\,a}$\vspace{0.1cm}} 
&  fund. \\
\hline
\end{tabular}
\end{center}
\caption{Summary of fields in the ADHM model.}
\label{tab2}
\end{table}

\noindent We have not listed fields coming from 5-5 and 5-9 strings,
since here we are only interested in the case of vanishing instanton
size which corresponds to setting the 5-9 fields to zero (see
\cite{Witten10, Lambert}).  Moreover, the 5-5 fields $X^{AY}_{mn}$
denote the position of the D5-branes in the transversal space, which
we treat as parameters of the low energy theory.\footnote{As we will
  explain in more detail when discussing the orbifolded theory in
  section \ref{secorbifold}, the D5-branes will all be clustered at
  the orbifold fixed point ($x^2=x^3=x^4=x^5=0$), which instructs us
  to set $X^{AY}_{mn}=0$.}\\

\noindent 
The Lagrangian describing the low-energy physics of the type~I
D1-D5-D9 intersection can now be written in terms of the fields of
table~\ref{tab2}. For $p \geq 1$, it is convenient to divide the
Lagrangian into three parts,
\begin{align} \label{flatspaceaction}
{\cal L} &= {\cal L}_{\rm kin} + {\cal L}_{\rm pot}
           + {\cal L}_{\rm int} \,,
\end{align}
where ${\cal L}_{\rm kin}$ contains the kinetic terms for all fields
in table~\ref{tab2}, and ${\cal L}_{\rm pot}$ describes their
potential. In general, ${\cal L}_{\rm pot}$ contains Yukawa couplings
of the type $b\psi_+\psi_-$ and D-terms for the scalars~$b$. For
details, see ref.~\cite{Lowe}.

\noindent
The Lagrangian ${\cal L}_{\rm int}$ describes the interaction of 1-1 
with 1-5 string modes and is given by \cite{Douglas, Lowe}
\begin{align}
{\cal L}_{\rm int}  &= {\rm Tr\,} \left( \frac{i m}{2} \big(\psi_-^{A'Y}
     \chi_{+\,Ym} + \psi_+^{A A'} \chi_{-\,A m} \big) \phi_{A'}{}^m
       + \frac{i m}{2} \chi_{+\,Ym} (X_{mn}^{AY}-b^{AY} \delta_{mn}) 
            \chi_-^{An} \right. \nonumber \\
 &  
\qquad\quad\left. + \frac{m^2}{8} (X_{mn}^{AY}-b^{AY} \delta_{mn})^2
            \phi_{A'm} \phi^{A'n}  \right) + {c.c.} \,,
\end{align}
where the trace is taken over the $SO(p)$ indices.  As first found in
\cite{Douglas} for $p=1$, this Lagrangian corresponds to Witten's ADHM
model \cite{Witten:1994tz} describing an $Sp(2N')$ instanton with
instanton number one. It is believed that for $p \geq 1$ the ADHM
model describes the moduli space of $Sp(2N')$ instantons with
instanton number $p$.

\subsection{ADHM orbifold theory}\label{secorbifold}

\subsubsection{Field content of the orbifold theory}
Let us now include the effect of the KK monopoles in the ADHM model.
This requires us to consider the D1-D5-D9 intersection at the origin
of a $\CC^2/\ZZ_N$ orbifold acting along $x^{2,3,4,5}$.  Following
Refs.~\cite{Douglas:1996sw, Johnson:1996py, Johnson}, we start with
$pN$ D1-branes intersecting $2 N' N$ D5-branes and $32N$ D9-branes in
flat space and take the corresponding ADHM Lagrangian with gauge group
$U(Np) \times U(2 NN') \times U(32N)$ as the parent
theory.\footnote{Formally, we begin with the type IIB version of the
  ADHM model \cite{Douglas} and perform the orientifold projection in
  the next subsection. The overall factor 2 in $U(2NN')$ reflects the
  pairing of the D5-branes for invariance under $\Omega$.} The ADHM
orbifold theory is then obtained by projecting out the degrees of
freedom which are not invariant under the $\ZZ_N$ orbifold group.

The $\CC^2/\ZZ_{N}$ orbifold is realized as follows. Denote the matrix
$b^{AY}$ by
\begin{align}
 b = (b^{AY}) = 
\begin{pmatrix}
b^1 & - \bar b^2\\
b^2 & \bar b^1 
\end{pmatrix} \,,
\end{align} 
where $b^1 = x^2+ix^3$ and $b^2 = x^4+ix^5$. Then the action of $(g_A,
g_Y) \in SO(4)_E=SU(2)_A \times SU(2)_Y$ along $x^{2,3,4,5}$ is
realized by
\begin{align}
b \mapsto g_Y b g_A \,.
\end{align}
We now embed the $\ZZ_{N}$ action in $SU(2)_Y$ by choosing $g_Y= {\rm
  diag} (\omega, \omega^{-1})$ with $\omega=e^{2\pi i/N}$.  Then,
\begin{align}
b^1 \mapsto \omega b^1\,, \quad b^2 \mapsto \omega^{-1} b^2 \,,
\end{align}
or, alternatively, $b^{AY}\mapsto \omega^Y b^{AY}$.  The scalars
$b^{A'\tilde A'}$ (along $x^{6,7,8,9}$) remain unaffected by the
orbifold.  The origin of $x^{2,3,4,5}$ is the only fixed point of the
orbifold.

The orbifold therefore acts on the fields of the ADHM model as follows
(gauge indices suppressed):
\begin{align}
1-1: &\qquad 
b^{AY} \rightarrow \omega^Y g_1(\omega) b^{AY} g^\dagger_1(\omega)\,,\qquad 
\psi_-^{A'Y} \rightarrow \omega^Y g_1(\omega) \psi_-^{A'Y} g^\dagger_1(\omega) 
\,, \nonumber\\
&\hspace{6.3cm} \psi^{Y\tilde A'}_{+} \rightarrow \omega^{Y}  
g_1(\omega) \psi^{Y\tilde A'}_{+} g_1^\dagger(\omega) \,, \nonumber\\
&\qquad
b^{A'\tilde A'} \rightarrow g_1(\omega) b^{A'\tilde A'} g^\dagger_1(\omega) 
\,, \qquad \,\,\,
\psi_-^{A\tilde A'} \rightarrow g_1(\omega) \psi_-^{A\tilde A'} g^\dagger_1(\omega) 
\,,\nonumber\\
&\hspace{6.3cm} \psi^{A'A}_{+} \rightarrow  g_1(\omega)  
\psi^{A'A}_{+}  g^\dagger_1(\omega) \nonumber\\
\nonumber\\
1-5:&\qquad 
\phi^{A'} \rightarrow g_1(\omega) \phi^{A'} g^\dagger_5(\omega)\,,\qquad 
\chi_-^{A} \rightarrow g_1(\omega) \chi_-^{A} g^\dagger_5(\omega)
\,,  \nonumber\\
&\qquad
\chi_+^{Y} \rightarrow \omega^Y g_1(\omega) \chi_+^{Y} g^\dagger_5(\omega) 
\, \nonumber\\\nonumber\\
1-9: &\qquad \lambda_+ \rightarrow g_1(\omega) \lambda_+ 
     g^\dagger_9(\omega) \label{orb0} \,.
\end{align} 
Here $g_1(\omega)$, $g_5(\omega)$, $g_9(\omega)$ denote the usual
embeddings of the $\ZZ_{N}$ orbifold group in the gauge groups
$U(Np)$, $U(2 NN')$ and $U(32N)$, respectively. We choose a basis such
that the embedding matrices have the block-diagonal form $g_i(\omega)
= {\rm diag}(\1, \omega \1, \omega^2 \1, ..., \omega^{N-1} \1)$, where
$\1$ denotes a $p\times p$, $2N'\times 2N'$ and $32\times 32$ unit
matrix for $i=1,5,9$, respectively. The fields thus decompose into $N$
orbifold sectors which we denote by $j,j'=0,\ldots, N-1$.  We observe
that all fields carrying an index $Y$ transform non-trivially under
the orbifold group, \ie the transformation law contains an additional
factor $\omega^Y$.

Substituting the embeddings $g_i(\omega)$ into (\ref{orb0}), we get
the following transformation behaviour in component form:
\begin{align}
1-1: &\qquad 
b^{AY}_{j,j'} \mapsto \omega^{Y+j-j'} b^{AY}_{j,j'}\, ,\qquad 
\psi_{-\,j,j'}^{A'Y} \mapsto \omega^{Y+j-j'} \psi_{-\,j,j'}^{A'Y}  
\,, \nonumber\\
&\hspace{5.2cm} \psi^{Y\tilde A'}_{+\,j,j'} \mapsto \omega^{Y+j-j'} 
\psi^{Y\tilde A'}_{+\,j,j'}\, \nonumber\\
&\qquad
b^{A'\tilde A'}_{j,j'} \mapsto \omega^{j-j'} b^{A'\tilde A'}  
\,, \qquad \,\,\,
\psi_{-\,j,j'}^{A\tilde A'} \mapsto \omega^{j-j'}
\psi_{-\,j,j'}^{A\tilde A'}   
\,,\nonumber\\
&\hspace{5.2cm} \psi^{A'A}_{+\,j,j'} \mapsto \omega^{j-j'} 
\psi^{A'A}_{+\,j,j'} \nonumber\\
\nonumber\\
1-5:&\qquad 
    \phi^{A'm}_{j,j'} \mapsto \omega^{j-j'} \phi^{A'm}_{j,j'} \,,\qquad 
    \chi_{-\,j,j'}^{Am} \mapsto \omega^{j-j'} \chi_{-\,j,j'}^{Am} 
    \,,  \nonumber\\
&\qquad
\chi_{+\,j,j'}^{Ym} \mapsto \omega^{Y+j-j'} \chi_{+\,j,j'}^{Ym}  
\, \nonumber\\\nonumber\\
1-9: &\qquad \lambda_{+\,j,j'}^M \mapsto \omega^{j-j'} 
     \lambda_{+\,j,j'}^{M} 
     \label{orb} \,
\end{align} 
where $Y=\pm 1$.\smallskip

\noindent
The fields invariant under the orbifold action (\ref{orb}) are thus
\begin{itemize}
\item 1-1: $(b^{A'\tilde A'}_{j,j}, \psi^{A\tilde A'}_{-\,j,j},
  \psi^{A'A}_{+\,j,j}$) and $(b^{A Y}_{j,j+Y},
  \psi^{A'Y}_{-\,j,j+Y},\psi^{Y\tilde A'}_{+\,j,j+Y})$
\item 1-5: $(\phi^{A'm}_{j,j}, \chi^{A m}_{-\, j,j},\chi^{Ym}_{+\, j,j+Y})$
\item 1-9: $\lambda_{+\,j,j}^M$
\end{itemize}
Another important question concerns the gauge groups and
representations under which these fields transform. Due to the
$\Omega$-projection of type~I string theory, this issue is more
intricate than in type~II theories and will now be discussed at length.

\subsubsection{Type I effective action}

The type~I effective theory is obtained by imposing, in addition
to the $\mathbb{Z}_N$ orbifold projection, the orientifold
$\Omega$~\cite{Douglas:1996sw}. Let us denote the embedding of
$\Omega$ into the gauge groups $U(Np)$, $U(2 NN')$ and $U(32N)$ by
$g_1(\Omega)$, $g_5(\Omega)$ and $g_9(\Omega)$, respectively. A
generic (scalar) field $y$ then transforms under worldsheet parity
according to \be y \mapsto g(\Omega) y^t g(\Omega)^{-1}\ ,
\label{matter} 
\ee 
while an element $U$ of one of the above gauge groups satisfies
\be 
U g(\Omega) U^t g(\Omega)^{-1} = 1\ .\label{gaugegroup} 
\ee 
Here $t$ denotes the transpose and $g$ is one of the embeddings $g_1,
g_5, g_9$.

To determine $g$, we have to solve various consistency
conditions~\cite{Douglas:1996sw}.  The first condition is
\begin{align}
&g(\Omega)_{ij} = \chi(\omega, \Omega) \omega^{i+j} g(\Omega)_{ij} \,.
\end{align} 
We choose the phase $\chi(\omega, \Omega) = 1$ which then implies that only
$g(\Omega)_{i,N-i}$ is non-vanishing.\footnote{For even orbifolds one
could also choose $\chi(\omega, \Omega)=\omega$, which would not invalidate our final conclusion. We will therefore not consider
this case here.}  A~second condition requires
\begin{align} \label{gO1}
g(\Omega)_{i,N-i} = \chi(\Omega) g(\Omega)^t_{N-i,i} \ , 
\end{align} 
with some phase factor $\chi(\Omega)=\pm 1$. To reproduce
the standard type~I action, which has an $SO(32)$ gauge
group for the D9-branes, we choose the phases $\chi(\Omega)=+1,-1,+1$
for $g=g_1, g_5, g_9$, respectively.\footnote{In fact, once we have
set $\chi_9(\Omega)=+1$, which is necessary to get a consistent
$SO(32)$ type I string theory, the other values follow (see
\cite{Gimon:1996rq}).} The solutions of
(\ref{gO1}) can be brought into the form
\begin{align}
g_{1,9}(\Omega)_{0,0}=\1 \,, &&g_5(\Omega)_{0,0}=\mathbf{\epsilon}\,, 
\nn\\
g_{1,9}(\Omega)_{i,N-i}=\1 \,, &&g_5(\Omega)_{i,N-i}=\1 \,, 
&& 0 < i < N/2 \,, \nn\\
g_{1,9}(\Omega)_{N-i,i}=\1 \,, & &g_5(\Omega)_{N-i,i}=-\1 \,, 
&& N/2 < i < N \,,
\label{solutiong}
\end{align}
where $\1$ is the corresponding $p\times p$, $2N'\times 2N'$ or
$32\times 32$ unit matrix.  For even orbifolds, we have in addition
\begin{align} 
g_{1,9}(\Omega)_{N/2,N/2}=\1 \,, \qquad 
g_5(\Omega)_{N/2,N/2}=\mathbf{\epsilon}\,. 
\nn
\end{align}

\medskip
Let us now determine
the unbroken gauge groups from
(\ref{gaugegroup}).  We distinguish between even and odd orbifolds:
\begin{itemize}
 \item \textbf{even $N$}\\
For $g=g_1$, the gauge group of the D1-branes is
\begin{align}
  G^1_{\rm even} 
&= \{ (U_0, U_1, ..., U_{N-1}): U_i U_{N-i}^t=1,
  0 \leq i \leq N \} \,
  \nonumber\\
  &= SO(p) \times U(p)^{N/2-1} \times SO(p) \, , \label{G1}
\end{align}
while for the D5-branes, it is
\begin{align}
G^5_{\rm even} 
&= \{ (U_0, U_1, ..., U_{N-1}): U_i U_{N-i}^t=1, 0 \leq i \leq N-1,
i \neq N/2 \} \, 
\nonumber\\
&= Sp(2N') \times U(2N')^{N/2-1} \times Sp(2N') \,. \label{G5}
\end{align}
\item\textbf{odd $N$}\\
For $g=g_1$, we get the gauge group
\begin{align}
  G^1_{\rm odd} 
&= \{ (U_0, U_1, ..., U_{N-1}): U_i U_{N-i}^t=1,\,
  1 \leq i \leq N-1 \} \,
  \nonumber\\
  &= SO(p) \times U(p)^{\frac{N-1}{2}} \,, 
\end{align}
while for $g=g_5$, it is
\begin{align}
G^5_{\rm odd} 
&= \{ (U_0, U_1, ..., U_{N-1}): U_i U_{N-i}^t=1,\, 1 \leq i \leq N-1 \} \, 
\nonumber\\
&= Sp(2N') \times U(2N')^{\frac{N-1}{2}}  \, .
\end{align}
\end{itemize}
The effect on the matter fields is as follows.  For the $b^{AY}$,
equation~(\ref{matter}) reads
\begin{align} 
(b^{AY}_{N-i-Y, N-i})^t = b^{AY}_{i,i+Y}\ . \label{condbY} 
\end{align} 
For $N$ even, this relates one half of the fields to the other half,
but gives no additional constraints. The same holds true for the
fermions $\psi^{Y\tilde A'}_{+}$ and $\psi^{A' Y}_-$.  If $N$ is odd,
there is the additional condition
\be
(b^{AY}_{(N-Y)/2,(N+Y)/2})^t = b^{AY}_{(N-Y)/2,(N+Y)/2}\ , \label{Noddspecb}
\ee
so that these particular $b$ transform in the symmetric instead of the
bifundamental. The situation is analogous to the analysis in
section~\ref{secADHM}, so that their fermionic partners
$\psi^{A'Y}_{-\,(N-Y)/2,(N+Y)/2}$ and $\psi^{Y\tilde
  A'}_{+\,(N-Y)/2,(N+Y)/2}$ transform in the symmetric and
antisymmetric, respectively.

The $b^{A'\tilde A'}$ are subject to
\begin{align}
(b^{A'\tilde A'}_{i,i})^t = b^{A'\tilde A'}_{N-i,N-i}
\end{align}
for all $i=0, \ldots, (N-1)/2$ for $N$ odd and $i=0, \ldots, N/2$ for
$N$ even. Note that the fields $b^{A'\tilde A'}_{00}$ (and also 
$b^{A'\tilde A'}_{N/2,N/2}$ if $N$ is
even) are symmetric.  Again,
the situation is exactly as described above such that the
corresponding fermionic modes, $\psi^{A\tilde A'}_{-\,0,0}$ and
$\psi^{A' A}_{+\,0,0}$ (and $\psi^{A\tilde A'}_{-\,N/2,N/2}$ and
$\psi^{A' A}_{+\,N/2,N/2}$ for $N$ even), are in the symmetric and
anti-symmetric representation, respectively.

We omit the corresponding relations for the $\phi^{A'm}_{i,i}$, as
they again only relate half of the fields to the other half
\cite{Douglas:1996sw}.

\subsubsection{Quiver theory}
\begin{figure}[ht]
\input{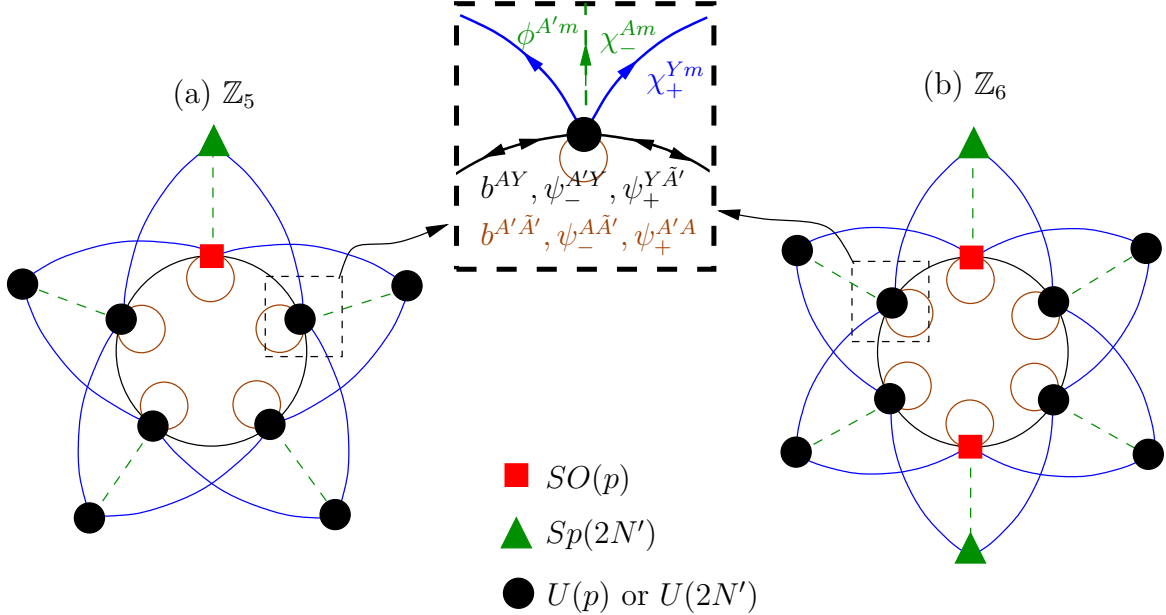}
\caption{Quiver diagrams for odd ($\mathbb{Z}_5$) and even
  ($\mathbb{Z}_6$) $N$. The detail view in the centre shows the
  notation for the fields. For simplicity, we have not included the
  fields $\lambda^M_+$.}
\label{quiverfig}
\end{figure}

So far we have determined the spectrum of fields that survive the
orientifold projection along with the gauge groups of the world-volume
theories of the various branes. It remains to determine the
representations under which the matter fields transform. In fact they
are given by
\begin{align}
b^{A'\tilde A'}_{j,j}, \psi^{A\tilde A'}_{-\,j,j}, \psi^{A'A}_{+\,j,j} &\qquad 
\textmd{adjoint rep.\ if\,\,} G^{1}_j=U(p) \,,\nonumber \\
&\qquad \textmd{rep.\ as in table~\ref{tab2} if\,\,} G^{1}_j=SO(p) \,,\nonumber \\
b^{A Y}_{j,j+Y}, \psi^{A'Y}_{-\,j,j+Y}, \psi^{Y\tilde A'}_{+\,j,j+Y}  &\qquad 
\textmd{bifundamentals of } G^{1}_j
\times G^{1}_{j+Y} 
\,,\nonumber\\ 
\phi^{A'm}_{j,j}, \chi^{A m}_{-\, j,j} &\qquad 
\textmd{bifundamentals of } G^{1}_j \times G^{5}_j\,, \nonumber\\
\chi^{Ym}_{+\, j,j+Y} &\qquad 
\textmd{bifundamentals of } G^{1}_j \times G^{5}_{j+Y} \, .\nonumber
\end{align}
The gauge groups and matter content of the theory can now be encoded
in a quiver diagram, see figure~\ref{quiverfig} for
examples.\footnote{A similar quiver diagram was also found in
  \cite{Constable:2002vt} for the $(0, 4)$ quiver theory located on a
  D3/D3$'$ intersection at a $\CC^2/\ZZ_N$ orbifold.}

\noindent
Each node in the inner circle corresponds to a gauge group $G^{1}_j$
(D1-branes), while an outer node represents a gauge group $G^{5}_j$
(D5-branes). In principle, there are also nodes corresponding to
$SO(32)$ gauge groups (D9-branes). The latter are not needed for the
interaction Lagrangian and are therefore not shown in
figure~\ref{quiverfig}. The fields $b^{A'\tilde A'}_{j,j}, \psi^{A\tilde
  A'}_{-\,j,j}$, $\psi^{A'A}_{+\,j,j}$ transform under a single gauge
group and are represented as brown circles. The bifundamentals
$b^{AY}_{j,j+Y}, \psi^{A'Y}_{-\, j,j+Y}$, $\psi^{Y\tilde
  A'}_{+\,j,j+Y}$ (shown as black lines), $\phi^{A'm}_{j,j}, \chi^{A
  m}_{-\, j,j}$ (green lines), and $\chi^{Ym}_{+\, j,j+Y}$ (blue
lines) connect different nodes.  We have omitted bifundamentals
connecting the outer nodes. These are generated by 5-5 strings which
decouple at low-energies, as already discussed earlier.\\

\noindent
We may now write down the corresponding quiver Lagrangian which
descends from the ADHM Lagrangian in flat space,
Eq.~(\ref{flatspaceaction}). Upon projecting out the degrees of
freedom which are not invariant under the orbifold, we obtain
\begin{align} \label{lagrangianorbifold} 
{\cal L}&= {\cal L}_{\text{kin, quiv}} + {\cal L}_{\text{pot, quiv}} 
+ {\cal L}_{\text{int, quiv}}
\end{align}
with the quiver interaction
\begin{align}
&{\cal L}_{\text{int, quiv}} = \nn\\
&\quad{\rm Tr\,} \left(\frac{i m}{2} (\chi_{+\,Ym})_{j,j+Y}
  (\phi_{A'}{}^m)_{j+Y,j+Y} (\psi_-^{A'Y})_{j+Y,j}\,+\frac{i m}{2} 
(\psi_{+}^{AA'})_{j,j} (\chi_{-\,A m})_{j,j} (\phi_{A'}{}^m)_{j,j}  \right.\nn\\
  &\quad \left. + \frac{i m}{2} (\chi_{+\,Ym})_{j,j+Y} (\chi_-^{Am})_{j+Y,j+Y}
  (b_{A}{}^Y)_{j+Y,j} + \frac{m^2}{8} (b^{AY}b_{AY})_{j,j} (\phi^{A'm}
  \phi_{A'm})_{j,j} \right) + {c.c.} \,, \nn
\end{align}
and, similarly, ${\cal L}_{\text{kin, quiv}}$ and ${\cal L}_{\text{pot,
    quiv}}$ are the projections of ${\cal L}_{\text{kin}}$ and ${\cal
  L}_{\text{pot}}$ in (\ref{flatspaceaction}), respectively. The range
of summation over $j$ and $Y$ is restricted by the $\Omega$
projection. For instance, for $N$ even, consider again the quiver
diagram shown in figure~\ref{quiverfig}. Each Yukawa coupling
corresponds to a triangle in the quiver diagram. The field
identifications of the previous section introduce a kind of reflection
axis, which vertically divides the quiver in two parts. The $SO(p)$
gauge groups at $j=0, N/2$ lie on the $\ZZ_2$ reflection axis. Due to constraints such as (\ref{condbY}), each field on the right hand side of
the axis is identified with one on the left hand side.  In
(\ref{lagrangianorbifold}) we therefore sum only over $j=0, ..., N/2$
and set $Y=+1$ at $j=0$ and $Y=-1$ at $j=N/2$, $Y=\pm 1$ otherwise.
The gauge groups are chosen as in (\ref{G1}) and (\ref{G5}). For $N$
odd, the Lagrangian is constructed in a similar way.


\subsection{Higgs branch theory and instanton moduli space}\label{secinst}
\subsubsection{Higgs branch theory}

In this section we investigate the infrared fixed point theory of the
ADHM quiver model (\ref{lagrangianorbifold}). This theory will be
interpreted as the boundary conformal field theory dual to the
worldsheet theory described in section~\ref{secws}. For its
construction, we first have to choose a vacuum solution which sets the
potential of (\ref{lagrangianorbifold}) to zero.  Inspecting the term
$m^2 b^2 \phi^2$ in (\ref{lagrangianorbifold}), we find two different
possibilities for the scalars $b^{AY}$ and $\phi^{A'm}$ and their
vacuum expectation values $\langle b^{AY}\rangle$ and
$\langle\phi^{A'm}\rangle$ \cite{Witten:1994tz}:\footnote{We will not
  discuss the rather delicate case $\langle
  b^{AY}\rangle=\langle\phi^{A'm}\rangle=0$.}
\begin{itemize}
\item Coulomb branch: 
$\langle b^{AY}\rangle\neq0$ and $\langle\phi^{A'm}\rangle= 0$\\
On the Coulomb branch the D1-branes are transversely displaced from
the D5-branes with $\langle b^{AY}\rangle$ proportional to the
distance. In this case the $\phi^{A'm}$ become massive.
\item Higgs branch: $\langle b^{AY}\rangle=0$ and 
$\langle\phi^{A'm}\rangle\neq 0$\\ 
On the Higgs branch the D1-branes and D5-branes form a bound state
with $\langle\phi^{A'm}\rangle$ proportional to the binding strength
between the two. In this case the $b^{AY}$ become massive.
\end{itemize}
In the following we are interested in the situation where all branes
form stacks located at the orbifold fixed point. We will therefore
consider the Higgs branch of the theory.

In principle, we could now proceed as in \cite{Lambert} and integrate
out all massive modes of the quiver theory. As in \cite{Lambert}, this
would lead to a $(0, 4)$ sigma model whose target space is the
instanton moduli space ${\cal M}$ of the ultraviolet theory. The
actual construction would be along the lines of \cite{Lambert} and
involves a non-trivial gauge field $F^{pq}_{mn\,jj}$ which is defined
in terms of the bifundamentals $\phi^{A'm}_{jj}$. Although
straightforward, we will not do this explicitly here. Instead we only
determine the left- and right-moving central charges of the infrared
theory and compare them to those expected from the dual worldsheet
model.

As outlined in the introduction, our strategy to find these charges is
as follows. The ADHM quiver model is classically not conformally
invariant, but ultraviolet finite such that there is no
renormalisation group flow.  This follows from the fact that the
one-loop diagrams cancel, and all higher loop diagrams are finite
\cite{Lambert2}. The massless fields of the quiver model therefore do
not acquire anomalous conformal dimensions and contribute to the
central charges of the infrared conformal field theory.  This allows
us to determine the left- and right-moving central charges of the
infrared conformal field theory from the number of massless modes in
the ultraviolet quiver theory.

\subsubsection{Number of massless modes for $N$ even}

We begin by counting the massless degrees of freedom in the case of
even orbifolds: First, there are the bifundamental fields
$(\phi^{A'm}_{a})_{j,j}$ descending from 1-5 strings and their left-
and right-moving fermionic partners $(\chi^{Am}_{-\,a})_{j,j}$ and
$(\chi^{Ym}_{+\,a})_{j,j+1}$. These fields are not constrained by any
D-term relations and thus contribute $2 \cdot N \cdot 2 N'\cdot p =
4NN'p$ scalars and an equal number of left- and right-moving
fermions. The 5-1 string modes are related to the 1-5 modes by the
$\Omega$ reflection and therefore do not contribute any additional
massless modes.

Second, consider the bosons $(b^{AY}_{ab})_{j,j+Y}$ which are massive
on the Higgs branch.  Since the theory has $(0,4)$-supersymmetry, we
know immediately that an equal number of right-moving fermions
$(\psi^{A'Y}_{-\,ab})_{j,j+Y}$ has to obtain mass. However, since only
non-chiral fermions can be massive, it follows that also all
left-moving $(\psi_{+\,ab}^{AA'})_{j,j}$ become massive. The mass
terms for the latter arise due to couplings of the type $\psi_+ \chi_-
\phi$ in (\ref{lagrangianorbifold}).  This sector thus has no
massless modes.

Third, consider the scalars $(b^{A'\tilde A'}_{ab})_{j,j}$. Those
fields $(b^{A'\tilde A'}_{ab})_{j,j}$ which are adjoints of a $U(p)$
gauge group do not contribute to the counting: The $4p^2$ degrees of
freedom of $(b^{A'\tilde A'}_{ab})_{j,j}$ (for fixed $j \neq 0, N/2$)
are removed by $3p^2+p^2$ conditions coming from the vanishing of the
corresponding D-term and $U(p)$ gauge equivalence.  By supersymmetry,
the same number of $(\psi^{A\tilde A'}_{-\,ab})_{j,j}$ are removed,
and by the same pairing mechanism as described above also all of the
$(\psi^{Y\tilde A'}_{+\,ab})_{j,j+Y}$. These fields thus give no
contribution.

For $j=0$ and $j=N/2$, however, the gauge group is $SO(p)$, and the
counting is similar as in the unorbifolded case
\cite{Douglas:1995bn, Barbon:1998nx}: the fields $(b^{A'\tilde
A'}_{(ab)})_{0,0}$ and $(b^{A'\tilde A'}_{(ab)})_{N/2,N/2}$ are in the
symmetric representation of $SO(p)$ and contribute $4 p(p+1)/2$ real
scalars each. However, there are also $4 p(p-1)/2$ constraints due to
D-term relations and gauge equivalences. In total, $(b^{A'\tilde
A'}_{(ab)})_{0,0}$ and $(b^{A'\tilde A'}_{(ab)})_{{N/2},{N/2}}$ thus
contribute $2(4 p(p+1)/2-4 p(p-1)/2) = 8p$ massless bosons.
Supersymmetry then dictates that of the $8 p(p+1)/2$ right-moving
fermions $(\psi^{A'\tilde A'}_{-(ab)})_{0,0}$ and $(\psi^{A'\tilde
A'}_{-(ab)})_{N/2,N/2}$ only $8p$ survive. To eliminate the remaining
$8p(p-1)/2$, we need to pair up all of the $8p(p-1)/2$ left-moving
fermions $(\psi^{A'\tilde A'}_{+[ab]})_{0,0}$ and $(\psi^{A'\tilde
A'}_{+[ab]})_{N/2,N/2}$. This leaves us with no left-moving massless
fermions.

\subsubsection{Number of massless modes for $N$ odd}
Much of the above analysis carries over to odd orbifolds. The fields
$(\phi_a^{A'm})_{j,j}$ again contribute $4N N' p$ massless bosonic
degrees of freedom and an equal number of left- and right-moving
fermions. For $j\neq 0$, the $(b_{ab}^{A'\tilde A'})_{j,j}$ of the
$U(p)$ gauge groups are eliminated by D-terms, and for $j=0$
$(b_{(ab)}^{A'\tilde A'})_{0,0}$ give $4p$ degrees of freedom. Note that
we only have one $SO(p)$ gauge group and we therefore get only half as
many massless degrees of freedom from these fields as required.

\begin{figure}[ht]
\begin{center}
\input{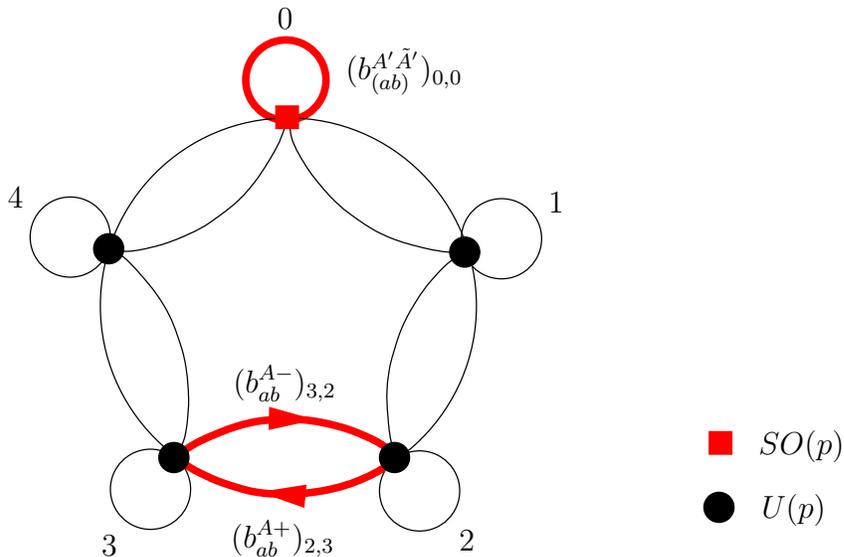}
\caption{Inner circle of the quiver diagram for an odd type~I orbifold
  ($\ZZ_{5}$). The special fields that contribute to the counting are
  denoted by red arrows.} 
\label{quiv2}
\end{center}
\end{figure}

Since we are on the Higgs branch, all the $(b_{ab}^{AY})_{j,j+Y}$
become massive, except for the fields
$(b_{ab}^{A+})_{(N-1)/2,(N+1)/2}$ and
$(b_{ab}^{A-})_{(N+1)/2,(N-1)/2}$ shown by red arrows in
figure~\ref{quiv2}. These fields are special and essentially take on
the role played by the second $SO(p)$ gauge group in the even case. By
(\ref{Noddspecb}) these particular $b^{AY}$ fields and their
superpartners $\psi_-$ are symmetric fields with $4p(p+1)/2$
components each, while the corresponding left-moving fermions $\psi_+$
are antisymmetric fields with $4p(p-1)/2$ components.  From the type
II theory we know that the only other left-moving fermions, the
$\chi_+$, remain massless. We can thus only form $4p(p-1)/2$ Yukawa
terms so that of the $\psi_-$, $4p(p+1)/2-4p(p-1)/2=4p$ remain. By
supersymmetry, the same number of bosons $b$ must remain massless. The
total number of bosonic degrees of freedom is thus again $4NN'p +8p$
(for $N>1$), the same as in the even case.

In the degenerate case $N=1$ there is one $SO(p)$ gauge group, but no
bifundamentals $b^{AY}$ of the type described above. We therefore get
only $4NN'p + 4p$ bosonic massless degrees of freedom, in agreement
with the unorbifolded ADHM model.

\subsubsection{Central charges of the Higgs branch theory}

From the above counting of massless degrees of freedom, we find that
the moduli space of the ultraviolet theory is spanned by the $4NN'p$
fields $(\phi^{A'm}_{a})_{j,j}$ and the $8p$ independent degrees of
freedom provided by $(b^{A'\tilde A'}_{(ab)})_{j,j}$ ($j=0,N/2$).  Its
dimension is therefore given by 
\begin{align} \label{dimensionM}
{\rm dim\,}{\cal M} = 4NN'p + 8p \,.
\end{align}
Recalling that the target space of the conformal sigma model on the
Higgs branch is the instanton moduli space of the ADHM quiver model,
we may now also determine the central charges of the infrared
theory. For $N \geq 2$ we find
\begin{align}
(c_L, c_R) 
  &=(6 N N' p + 24 p, 6 N N' p+ 12 p) \,
\end{align}
in agreement with (\ref{cc1}) and (\ref{cc2}).  The leading term,
$6NN'p$, is given by the ADHM instanton fields $\phi^{A'm}_{jj}$ and
their fermionic partners (\mbox{1-5} strings). The subleading term in
the right sector, $12p$, is given by the conformal charges of the $8p$
independent degrees of freedom of the scalars $b^{A'\tilde A'}_{jj}$
and their fermionic superpartners (1-1 strings). One contribution to
the term $24p$ in the left-moving sector is given by the $8p$ bosonic
fields descending from the $b^{A'\tilde A'}_{jj}$.  The remaining
$16p$ are given by the 32 fermions $\lambda^M_{+a}$ (1-9 strings).\\

\noindent
In conclusion, we propose the $(0, 4)$ sigma model on the Higgs branch
of the type~I quiver model (\ref{lagrangianorbifold}) as the
appropriate candidate for the boundary conformal field theory of
heterotic string theory on $AdS_3 \times S^3/\ZZ_{N} \times T^4$ ($N
\geq 2$).

\setcounter{equation}{0}
\section{Entropy function formalism in 5-dimensional heterotic 
string theory}\label{EntropyFunction}

\subsection{Outline}\label{classificobjects}
In this section we return to the construction of the near-horizon
geometry of the heterotic three-charge model. The corresponding
classical supergravity solution has been reviewed in
section~\ref{sec21}.  We now wish to go beyond classical 
supergravity by
introducing additional higher-derivative operators in the heterotic
string action. 
Our calculations are valid only for large values of the charges 
$N, N', q$. In particular, we shall only calculate the first
subleading correction to the classical solution.

Similar computations in a dual setup have already been performed in
\cite{CDKL} which exploit the recently discovered $\N=2$ off-shell
completion of the $R^2$-terms in the 5-dimensional supergravity action
\cite{Hanaki:2006pj}. Here, we will study the modification of the
near-horizon solution (\ref{AdSmetric}) in the presence of the four
derivative corrections to the heterotic string effective action
at the string tree level \cite{Metsaev:1987zx, Hull:1987pc}.

We will make use of the entropy function formalism; for an
introduction see e.g.~\cite{Sen:2005wa,Sen:2005iz} or the recent
review \cite{Sen:2007qy}. It was originally developed for
4-dimensional $AdS_2\times S^2$ black holes, but it can also be
generalised to geometries containing $AdS_p$-factors with $p>2$ (see
e.g.~\cite{Garousi:2007zb}). We will first use the formalism to
rederive the classical contribution to the central charge. In a second
step we then apply it to the $\alpha'$ corrected action to obtain
corrections to the central charge.  

Generically, the 5-dimensional action will also contain Chern-Simons
like contributions which contain the gauge fields in a non-covariant
way ({\em i.e.}\ terms which contain the gauge potentials rather than
the field strengths). We therefore cannot use the entropy function
formalism in a straightforward way. Fortunately, following
\cite{Sahoo:2006pm}, we can circumvent this problem by considering the
theory in 6 dimensions, from where we can get the 5-dimensional theory
by Kaluza-Klein reduction. This approach has not only the advantage
that we can reformulate the gauge Chern-Simons term in a covariant
way, but it also allows us to think of the 5-dimensional 2- and 3-form
field strengths as coming from the Kaluza-Klein reduction of a single
6-dimensional three-form.  Since the latter is in fact self-dual, this
provides us with a very compact way of dealing with the 5-dimensional
fields.  We will see however that the action still contains a
gravitational Chern-Simons term which will require special
treatment.\\ 

\noindent
Throughout this section we will
use the convention $\alpha'=16$.


\subsection{Uncorrected solution}\label{Soluncorrect}

\noindent
We begin by lifting the heterotic theory to 6 dimensions, where we
have the following massless bosonic fields
\begin{itemize}
\item 6-dimensional metric $G^{(6)}_{MN}$:\\
  This reduces to the 5-dimensional metric as well as to a vector
  field under which the black string can be electrically and magnetically
  charged.
\item anti-symmetric tensor $B_{MN}^{(6)}$:\\
  This reduces to a 5-dimensional 2-form potential and to a dual
  vector field. The black string can be electrically and magnetically
  charged under $B_{MN}^{(6)}$.
\item 6-dimensional dilaton:\\
This reduces to the 5-dimensional dilaton.
\end{itemize} 
 
\noindent
In the following the convention for the indices will be
\begin{align}
&M,N\in 0,1,\ldots,5\,, &&\text{and} &&\mu,\nu\in 0,1,\ldots,4\,.\nonumber
\end{align}
The 6-dimensional Lagrangian obtained from heterotic string theory 
is given by
\begin{align}
\mathcal{L}^{(6)}=\frac{1}{32\pi}e^{-2\Phi^{(6)}}\left[R^{(6)} + 
4\partial_M\Phi^{(6)}\partial^M\Phi^{(6)}
-\frac{1}{12}H^{(6)}_{MNP}H^{(6),MNP}\right]\, . \label{heteroticLagrangian}   
\end{align}
The 3-form field strength is given by
\begin{align}
H^{(6)}_{MNP}=\partial_MB^{(6)}_{NP}+\partial_NB^{(6)}_{PM}
+\partial_PB^{(6)}_{MN}+\kappa\Omega^{(6)}_{MNP}\,,\nonumber
\end{align}
where $\Omega^{(6)}_{MNP}$ is the gravitational Chern-Simons 3-form.
The parameter $\kappa$ can be fixed as in \cite{Sahoo:2006pm}, which
gives the value $\kappa=192$ for our setup.
To covariantise the action, we 
introduce a new field $\ck_{MN}$ together with its field strength
\begin{align}
\mk_{MNP}=\partial_M\ck_{NP}+\partial_N\ck_{PM}+\partial_P\ck_{MN}\,.
\label{Kfieldstrength}
\end{align}
Consider the new Lagrangian
\begin{align}
\mathcal{L}^{(6)}_{[1]}=\, &
\frac{\sqrt{-\text{det}G^{(6)}}}{32\pi}e^{-2\Phi^{(6)}}\left[R^{(6)}+
  4\partial_M\Phi^{(6)}\partial^M\Phi^{(6)}-
  \frac{1}{12}H^{(6)}_{MNP}H^{(6),MNP}\right]\nonumber\\ 
&+\zeta\epsilon^{MNPQRS}\mk_{MNP}
H^{(6)}_{QRS}-\zeta\kappa\epsilon^{MNPQRS}\mk_{MNP}\Omega^{(6)}_{QRS}\,,
\label{hetactfirstextens} 
\end{align}
where $\zeta$ is some constant which will cancel out in all physical
quantities.  Upon exploiting the equations of motion for the auxiliary
field $C^{(6)}$,
\begin{align}
\zeta\epsilon^{MNPQRS} \partial_P(H^{(6)}_{QRS}-\kappa \Omega^{(6)}_{QRS})=0
\,,\nonumber
\end{align}
this reduces (\ref{hetactfirstextens})
to the old
Lagrangian (\ref{heteroticLagrangian}).
On the other hand we can use the
equation of motion for $H^{(6)}_{MNP}$
to get
\begin{align}
H^{(6),MNP}= -\frac{192\pi e^{2\Phi^{(6)}}}{\sqrt{-\text{det}G^{(6)}}}
\zeta\epsilon^{MNPQRS}\mk_{QRS}\,,\label{solhclass} 
\end{align} 
which we use to eliminate $H^{(6),MNP}$ from the original Lagrangian
(\ref{heteroticLagrangian}). 
We have thus replaced the 3-form field strength of the
6-dimensional Lagrangian by the (auxiliary-)field $\ck_{MN}$, which
only appears through its field strength $\mk_{MNP}$.

Let us comment briefly on the gravitational
Chern-Simons term
\be \label{CS}
-\zeta\kappa\epsilon^{MNPQRS}\mk_{MNP}\Omega^{(6)}_{QRS}\, .
\ee
Although it is not of a manifestly covariant form,  
we will argue below that in
our specific setup the term is actually
covariant.
This means that after replacing $H^{(6),MNP}$ by $\mk_{MNP}$,
(\ref{hetactfirstextens}) is covariant, so that we can
apply the entropy function formalism.

Although it will be more convenient to stay
in the 6-dimensional setup, let us spell out the ansatz
with which we can reduce this Lagrangian back to 5 dimensions:
\begin{align}
&\hat{G}_{55}=G^{(6)}_{55}\,, &&\hat{C}_{55}=C^{(6)}_{55}=0\,,\nonumber\\
&\hat{G}^{55}=(\hat{G}^{-1})^{55}\,, &&
G_{\mu\nu}=G^{(6)}_{\mu\nu}-
\hat{G}^{55}G^{(6)}_{5\mu}G^{(6)}_{5\nu}\,,\nonumber\\  
&A^{(1)}_{\mu}=\frac{1}{2}\hat{G}^{55}G^{(6)}_{5\mu}\,, &&
A^{(2)}_{\mu}=\frac{1}{2}C_{5\mu}\,,\nonumber\\ 
&C_{\mu\nu}=C^{(6)}_{\mu\nu}-2(A^{(1)}_{\mu}A^{(2)}_\nu -
A^{(1)}_{\nu}A^{(2)}_\mu)\,, && \Phi=\Phi^{(6)}-\frac{1}{2}\ln V\,,
\label{5dconfig}
\end{align}
where $V$ is the volume of the compactified $x_5$-direction. The field
strengths of the various forms are then given by
\begin{align}
&F^{(i)}_{\mu\nu}=\partial_\mu A^{(i)}_\nu-\partial_\nu A^{(i)}_{\mu}\,, 
\hspace{2cm}i=1,2\,, \nn\\
&\mathcal{K}_{\mu\nu\rho}=(\partial_\mu C_{\nu\rho}+2A^{(1)}_\mu 
F^{(2)}_{\nu\rho}+2A^{(2)}_\mu F^{(1)}_{\nu\rho})
+\text{cyclic permutation of }(\mu,\nu,\rho)\,.\label{3-formfieldstrength}
\end{align}
Note that after compactifying to 5 dimensions,
$\mathcal{K}_{\mu\nu\rho}$ is no longer covariant, as it contains
$A^{(1,2)}_{\mu}$ explicitly.  In principle, one would therefore have
to introduce new auxiliary fields and repeat the steps performed above
(see \cite{Sahoo:2006pm}). It turns out however that this gives the
same result as when we use the reduced version of
(\ref{hetactfirstextens}) directly.

We are now in a position to compute the entropy function, which is
given by
\begin{eqnarray}
\mathcal{E}_0 & = & \frac{2\pi}{r}\bigg\{q_ie_ir-\int_{\theta,\varphi,x_5}
\bigg[\frac{\sqrt{-\text{det}G^{(6)}}e^{-2\Phi^{(6)}}}{32\pi}
\bigg(R^{(6)}+4\partial_M\Phi^{(6)}\partial^M\Phi^{(6)}\nonumber\\
&& -\frac{1}{12}H^{(6)}_{MNP}H^{(6),MNP}\bigg)+ \zeta\epsilon^{MNPQRS}\mk_{MNP}H^{(6)}_{QRS}\nonumber\\
&& -\zeta\kappa\epsilon^{MNPQRS}\mk_{MNP}\Omega^{(6)}_{QRS}
 \bigg]\bigg\},\label{entropgeneralformclass}
\end{eqnarray}
where $H^{(6),MNP}$ is to be replaced by $\mk_{MNP}$ using
(\ref{solhclass}).
In order to evaluate (\ref{entropgeneralformclass}) we make the
following ansatz for the near-horizon form of all the 5-dimensional
fields involved
\begin{align}
&ds^2=g_{\mu\nu}dx^\mu dx^\nu=v_1(-r^2dt^2+r^2dz^2+\frac{dr^2}{r^2})+v_2(d\theta^2+\sin^2\theta d\varphi^2),\nonumber
\end{align}
\vspace{-0.85cm}
\begin{align}
&\hat{G}_{55}=u^2, && \nonumber\\
&F^{(1)}_{\theta\varphi}=\frac{p_1\sin\theta}{4\pi}, && F^{(2)}_{\theta\varphi}=-\frac{p_2\sin\theta}{4\pi},\nonumber\\
&F_{tr}^{(1)}=e_1, &&e^{-2\Phi}=\lambda \,, \label{5dansatz}
\end{align}
where we interpret $p_1$ and $p_2$ as magnetic and $e_1$ as the
Legendre transform of an electric charge. Using (\ref{5dconfig}), this
corresponds to the 6-dimensional configuration.
\begin{align}
&G^{(6)}_{MN} = \left(\begin{array}{cc}g_{\mu\nu} + u^2A_{\mu}A_{\nu}
    & u^2A_\mu \\ u^2 A_\nu & u^2\end{array}\right), &&\text{with }
A_\mu=\left\{\begin{array}{ccl} - \frac{p_2\cos\theta}{2\pi} & &
    \mu=\varphi \\ 0 & & \text{else}\end{array}\right. \,, \nn \\
&C^{(6)}_{tz}=2e_1r\,, &&C^{(6)}_{5\varphi}=\frac{p_1}{4\pi}\cos\theta\,,
\nn\\
&e^{-2\Phi^{(6)}}=\frac{\lambda}{u}\,.\label{ansatzentr3}
\end{align}
Let us now turn to the gravitational Chern-Simons term (\ref{CS}).  We
will argue that in our setup it is already covariant.  First note that
the 6-dimensional space factorizes into two 3-dimensional spaces,
which we label in the following way
\begin{align}
&\alpha,\beta,\gamma=t,r,z, &&\text{and} &&a,b,c=\theta,\varphi,x_5,\nonumber
\end{align}
where the metrics of the two subspaces read
\begin{align}
G_{\alpha,\beta}=v_1\left(\begin{array}{ccc}-r^2 & 0 & 0 \\ 0 &
    \frac{1}{r^2} & 0 \\ 0 & 0 & r^2\end{array}\right),
&&G_{\alpha,\beta}=\left(\begin{array}{ccc}v_2 & 0 & 0 \\ 0 &
    v_2\sin^2\theta+\frac{p_2^2u^2\cos^2\theta}{4\pi^2} &
    -\frac{p_2u^2\cos\theta}{2\pi} \\ 0 &
    -\frac{p_2u^2\cos\theta}{2\pi} & u^2\end{array}\right). 
\end{align}
The situation is now almost exactly as in \cite{Sahoo:2006pm}. There,
the setup was reduced to a two-dimensional geometry in $t,r$, since
all other directions were periodic and could thus be considered as
compactified. In our case although the $z$ direction is non-compact,
it does not appear explicitly in any of the expressions, so that the
argument carries over. The conclusion is then that (up to total
derivative terms which give no contribution) (\ref{CS}) is already
covariant, as was shown in \cite{Sahoo:2006pm}.

We can thus directly plug the expression for $\Omega^{(6)}$,
\be
\Omega^{(6)}_{MNP}=\frac{1}{2}\Gamma^R_{MS}\partial_N\Gamma^S_{PR}+\frac{1}{3}\Gamma^R_{MS}\Gamma^S_{NT}\Gamma^T_{PR}\, ,
\label{expressionCS}
\ee
into (\ref{CS}) to obtain the contribution
\be
\Delta\mathcal{E}_{\text{CS}}=-\frac{6 e_1 p_2 u^2 \left(p_2^2 u^2-4 \pi ^2 v_2\right) \zeta  \kappa }{\pi  v_2^2}\, \label{contCS}
\ee
with $\kappa=192$. A direct calculation shows however that 
$\Delta\mathcal{E}_{\text{CS}}$ only gives subleading corrections
to the classical geometry. We will thus omit the
Chern-Simons term as long as we consider the classical solution.

Inserting the ansatz (\ref{ansatzentr3}) into the entropy function
(\ref{entropgeneralformclass}), we obtain the result
\begin{align}
\mathcal{E}_0=&2 e_1  \pi  q_1-\frac{1}{2} \pi  v_1 ^{3/2} \lambda
+\frac{3}{2} \pi  \sqrt{v_1 } v_2  \lambda+\frac{p_2^2 v_1 ^{3/2}
  \lambda  u^2}{32 \pi  v_2 }-\frac{663552 e_1 ^2 \pi ^3 v_2  \zeta ^2
  u^2}{v_1 ^{3/2} 
\lambda }
+\frac{10368 p_1^2 \pi  v_1 ^{3/2} \zeta ^2}{v_2  \lambda }.
\end{align}

In order to find the entropy of the black hole, we have to
extremise this expression. Under the assumption $q_1>0$, $p_1>0$ and
$p_2>0$, the only physically acceptable extremum is
\begin{align}
&v_1=\frac{q_1p_2}{144\pi^2\zeta}\,, 
&&v_2=\frac{q_1p_2}{576\pi^2\zeta}\,, 
&&\lambda=\frac{6912p_1\pi\zeta^{\frac{3}{2}}}{\sqrt{q_1p_2}}\,, 
&&u=\frac{\sqrt{q_1}}{12\sqrt{p_2\zeta}} \,.\label{sol0}
\end{align}
Note in particular that we find the relation
\begin{align}
v_1=4v_2 \,.
\end{align}
We note, however, that the quantities $p_1,p_2,q_1$ are not yet
physically normalised expressions. The unphysical quantity $\zeta$
still enters into the solution. We will determine the correct normalisation
at the end of the next subsection.


\subsection{Corrected solution}
We now wish to consider corrections to the classical supergravity
theory.  This means that we have to include the
contribution (\ref{contCS}) of the Chern-Simons term.  Moreover,
the $\alpha'$-corrected supergravity Lagrangian also contains higher
order derivative terms which we have to take into account.
We follow \cite{Sahoo:2006pm} and
write down the action containing the four derivative corrections to
the heterotic string effective action as 
\begin{align}
\mathcal{L}^{(6)} & =
\frac{e^{-2\Phi^{(6)}}\sqrt{-\text{det}G^{(6)}}}{32\pi}\bigg[R^{(6)} +
4\partial_M\Phi^{(6)}\partial^M\Phi^{(6)}-
\frac{1}{12}H^{(6)}_{MNP}H^{(6),MNP} \nonumber\\
&+2R^{(6)}_{KLMN}R^{(6),KLMN}-R^{(6)}_{KLMN}H_P^{(6)KL}H^{(6),PMN} -
\frac{1}{4}H_{K}^{(6),MN}H^{(6)}_{LMN}H^{(6),KPQ}H_{PQ}^{(6),L}
\nonumber\\ 
&+\frac{1}{12}H^{(6)}_{KLM}H_{PQ}^{(6),K}H_R^{(6),LP}H^{(6),RMQ}\bigg]. 
\end{align}
As in the classical case, we introduce the new field
$C^{(6)}_{MN}$ with field strength $\mk_{MNP}$, as
defined in (\ref{Kfieldstrength}). As before, we modify
the action
\begin{align}
\mathcal{L}^{(6)}_{[1]} &
=\frac{e^{-2\Phi^{(6)}}\sqrt{-\text{det}G^{(6)}}}{32\pi}\bigg[R^{(6)}
+ 4\partial_M\Phi^{(6)}\partial^M\Phi^{(6)} -
\frac{1}{12}H^{(6)}_{MNP}H^{(6),MNP} \nonumber\\ 
&+ 2R^{(6)}_{KLMN}R^{(6),KLMN}-R^{(6)}_{KLMN}H_P^{(6)KL}H^{(6),PMN} -
\frac{1}{4}H_{K}^{(6),MN}H^{(6)}_{LMN}H^{(6),KPQ}H_{PQ}^{(6),L}
\nonumber\\ 
&
+\frac{1}{12}H^{(6)}_{KLM}H_{PQ}^{(6),K}H_R^{(6),LP}H^{(6),RMQ}\bigg]
+ \zeta\epsilon^{MNPQRS}\mk_{MNP} H^{(6)}_{QRS} \nonumber\\
&-\zeta\kappa\epsilon^{MNPQRS}\mk_{MNP}\Omega^{(6)}_{QRS} \,.
\end{align}
Reducing this Lagrangian to $\mathcal{L}^{(6)}$
by using the equations of motion for $C^{(6)}_{MN}$ is essentially the
same as in the classical case. However, elimination of $H^{(6)}_{MNP}$
is now modified due to the presence of the higher derivative
terms. Indeed, the equation of motion for $H^{(6)}_{MNP}$ now reads
\begin{align}
&\frac{\sqrt{-\text{det}G^{(6)}}}{32\pi} \bigg[-\frac{1}{6}H^{(6),MNP}
- 2{H^{(6),M}}_{KL}R^{(6),KLNP}
- \frac{1}{4}({H_L^{(6),NP}}H^{(6),MQR}{H^{(6),L}_{QR}} \nonumber\\
&+ H_K^{(6),NP}H^{(6),KQR}H_{QR}^{(6),M} +
H^{(6),MQR}H^{(6)}_{LQR}H^{(6),NPL} +
H_K^{(6),QR}{H^{(6),P}}_{QR}H^{(6),KMN}) \nonumber\\ 
&+ \frac{1}{12}(H_{QR}^{(6),M}H_K^{(6),NQ}H^{(6),KPR} +
{H^{(6),P}}_{LK}H^{(6),LM}_RH^{(6),RKN} +
H_{K\hspace{0.5cm}L}^{(6),N}H^{(6),P\hspace{0.3cm}K}_{\hspace{0.7cm}Q}
H^{(6),MLQ} \nonumber\\ 
&+H_{KL}^{(6),N}H^{(6),PK}_RH^{(6),MLR})\bigg] +
\zeta\epsilon^{MNPQRS}\mathcal{K}^{(6)}_{QRS}=0 \,. \label{eomfullh} 
\end{align}
Following the classical example, we would now have to invert this
equation to express $H^{(6)}_{MNP}$ in terms of
$\mathcal{K}^{(6)}_{MNP}$. Since this is in general very hard, we will
solve (\ref{eomfullh}) only to first subleading order.  To this end we
make the ansatz
\begin{align}
H^{(6),MNP}=H_0^{(6),MNP}+H_1^{(6),MNP} \,,
\end{align}
where $H_0^{(6),MNP}$ is the solution from the classical equations of
motion (see (\ref{solhclass})).  $H_1^{(6),MNP}$ is then a correction
to the classical solution, which is subleading in the charges.
Inserting this ansatz into (\ref{eomfullh}) and keeping only the first
subleading terms, we find the approximated solution
\begin{align}
H_1^{(6),MNP}= &-12H_{0\hspace{0.6cm}KL}^{(6),M}R^{(6),KLNP} -
\frac{3}{2}\bigg(3H_0^{(6),MQR}H^{(6)}_{0,QRL}H_0^{(6),LNP} \nonumber\\ 
&+H_0^{(6),MNK}H^{(6)}_{0,KQR}H_0^{(6),QRP}\bigg) +
2 H_0^{(6),MLQ}H_{0\hspace{0.6cm}KL}^{(6),N}
H_{0\hspace{0.7cm}Q}^{(6),PK}   
\,,\label{Hcorrect} 
\end{align}
where the right hand side is suitably antisymmetrised in
$M,N,P$. Indeed, we can justify 
our ansatz by plugging the classical solution 
into our results, to find
\begin{align}
&H_0^{(6),MNP}\sim\mathcal{O}(\text{charges}^{-4})
&&\text{and}&&H_1^{(6),MNP}\sim\mathcal{O}(\text{charges}^{-6})
\,.\nonumber 
\end{align}
This analysis makes it also clear that we only need to consider the
correction terms $H_1^{(6),MNP}$ in the classical terms, and not in
the higher derivative terms, where they only give sub-subleading
contributions.  The remaining steps of the preparation of the action
follow in exactly the same manner as for the classical case and can
therefore be literally carried over.


\medskip
\noindent
Now we are ready to compute the entropy function. Using 
(\ref{ansatzentr3}), (\ref{Hcorrect}), and (\ref{expressionCS}),
we find the following entropy function
\begin{multline}
\mathcal{E}=2 e_1 \pi q_1 -\frac{1}{2} \pi  v_1^{3/2} \lambda 
    +\frac{3}{2} \pi  \sqrt{v_1} v_2 \lambda     
    +\frac{p_2^2 v_1^{3/2} \lambda  u^2}{32 \pi  v_2} 
   - \frac{663552 e_1^2 \pi ^3 v_2 \zeta ^2 u^2}{v_1^{3/2} \lambda }\\
   +\frac{10368 p_1^2 \pi  v_1^{3/2} \zeta ^2}{v_2 \lambda } 
-\frac{1152 e_1 p_2^3 \zeta  u^4}{\pi  v_2^2} 
- \frac{11 p_2^4 v_1^{3/2} \lambda  u^4}{128\pi ^3 v_2^3} 
-\frac{331776 e_1^2 p_2^2 \pi  \zeta ^2 u^4}{v_1^{3/2} v_2 \lambda} \\ 
+ \frac{17612050268160 e_1^4 \pi ^5 v_2 \zeta ^4 u^4}{v_1^{9/2} \lambda ^3} 
    +\frac{4608 e_1 p_2 \pi  \zeta  u^2}{v_2} 
    +\frac{3 p_2^2 v_1^{3/2} \lambda  u^2}{4 \pi  v_2^2} 
    + \frac{5308416 e_1^2 \pi ^3 \zeta ^2 u^2}{v_1^{3/2} \lambda }  \\
    -\frac{6 \pi  v_2 \lambda }{\sqrt{v_1}} 
    -\frac{2 \pi  v_1^{3/2} \lambda }{v_2} 
   +\frac{248832 p_1^2 \pi  \sqrt{v_1} \zeta ^2}{v_2
   \lambda } + \frac{4299816960 p_1^4 \pi  v_1^{3/2} \zeta ^4}{v_2^3
   \lambda ^3} \,. 
\end{multline}

\noindent Since we are only interested in the first 
subleading correction,
we linearise around the uncorrected solution (\ref{sol0}) using
the ansatz
\begin{align}
&v_1=\frac{q_1p_2}{144\pi^2\zeta}+x_1\,, 
&&v_2=\frac{q_1p_2}{576\pi^2\zeta}+x_2\,, 
&&\lambda=\frac{6912p_1\pi\zeta^{\frac{3}{2}}}{\sqrt{q_1p_2}}+x_\lambda\,,
\nonumber\\
&u=\frac{\sqrt{q_1}}{12\sqrt{p_2\zeta}}+x_u\,, 
&&e_1=\frac{p_1p_2}{2\pi^2}+x_{e_1} \,.\label{sol1corr}
\end{align}
Extremising
$\mathcal{E}$ with respect to $(x_1, x_2, x_\lambda$, $x_u, x_{e_1})$
gives the first subleading terms as 
\begin{align}
&x_1=0\,, &&x_2=0\,, 
&&x_\lambda=31850496 \frac{\pi^3 p_1 \zeta^{5/2}}{p_2^{3/2}q_1^{3/2}}   \,,
 &&x_u=576 \frac{\pi ^2\zeta^{1/2}}{q_1^{1/2}p_2^{3/2}} \,, 
 &&x_{e_1}=0\,.
\end{align}

\noindent Let us finally normalise the charges $q_1, p_1, p_2$ and
relate them to the physical quantities $N',p,N$.
Following
\cite{Sahoo:2006pm}
we are led to identify
\begin{align}
&q_1=576\pi\zeta N',&&p_1=\frac{p}{144\zeta}, &&p_2=4\pi N.
\end{align}

\noindent To first order, the solution is then given by ($\alpha'=16$)
\begin{align} \label{correctedsol}
&v_1=16N N', &&v_2=4N N', 
&&\lambda=(N N')^{-1/2}p\left(1+\frac{2}{N N'}\right)\,,\nonumber\\
&u=\sqrt{\frac{N'}{N}}\left(1+\frac{3}{N N'}\right)  \,.
\end{align}
Note that the corrected solution still obeys
$v_1=4 v_2$. 

In summary, the corrected ten-dimensional near-horizon geometry is
still $AdS_3 \times S^3/\ZZ_N \times T^4$, but now with AdS radius and
six-dimensional string-coupling given by
\begin{align} 
R_{AdS,\, {\rm corr}}^2 = \alpha' NN'
+\mathcal{O}\left(\textstyle\frac{1}{NN'}\right)\,,\qquad  
g_{6,\, {\rm corr}}^{2}=\frac{u}{\lambda} = \frac{N'}{p} 
\left(1+ \frac{1}{NN'}+\mathcal{O}\left(\textstyle\frac{1}{(NN')^2}\right) \right) \,,
\end{align}
where it is understood that the sub-subleading terms can also be suppressed
by powers of $p$.
The Brown-Henneaux formula
\begin{align}
c=\frac{3}{8}\sqrt{v_1}v_2\lambda\,,
\end{align}
gives in the uncorrected case
\be
c_{\text{class}}=6NN' p\, ,
\ee 
while for the corrected solution we find
\begin{align} \label{cccorr}
c_{\text{corr}}=6NN'p + 12p+\mathcal{O}\left(\textstyle\frac{1}{NN'}\right)\,.
\end{align}
To subleading order this agrees with (\ref{cc1}).

\setcounter{equation}{0}
\section{Heterotic two-charge models}

In view of a possible heterotic string duality with $(0, 8)$ spacetime
supersymmetry \cite{Strominger, Kraus}, it is an interesting question
whether we can systematically switch off charges in the present $(0,
4)$ duality. Clearly, the worldsheet theory for strings on $AdS_3
\times S^3/\ZZ_N \times T^4$ requires at least one KK monopole and is
not applicable for vanishing KK monopole charge. Since the KK
monopoles break supersymmetry down to $(0, 4)$ there seems to be no
obvious way to generalise the model to $(0, 8)$. Nevertheless, it is
interesting to consider models with less charges such as the F1-KKM
and the NS5-KKM intersection.

\subsection{F1-KKM intersection (\texorpdfstring{$N'=0$}{})}

We shall first consider a heterotic two-charge model consisting of a
stack of $p$ fundamental strings in the background of a KK monopole
with charge proportional to $N \geq 2$. The setup is the same as in
section~2.1, but now $N'=0$ (no NS5-branes). From (\ref{cc1}), we find
the central charges of the boundary conformal field theory to be
$(c_L, c_R)=(24p, 12 p)$.  Remarkably, the central charges do not
depend on the charge of the KK monopole since the leading term cubic
in the charges ($\propto NN'p$) is absent. This has some interesting
consequences.

Let us first have a look at the supergravity solution. Classically,
the solution has a horizon of zero area leaving a naked curvature
singularity at the origin. This corresponds to a vanishing
Bekenstein-Hawking entropy on the classical level.  It is however
believed that higher-derivative corrections to the supergravity
solution resolve the classical singularity leading to a finite
entropy. The corrected supergravity solution presented in the previous
section is valid for large $NN'$ and thus cannot be applied to this
case.

The heterotic worldsheet theory for this case has some peculiar
features. The left sector of the CFT on the $S^3/\ZZ_{N}$ has
collapsed to a trivial theory with bosonic level $k_b'=c^{\rm
  ws}_L(S^3/\ZZ_{N})=0$.  The supersymmetric level corresponding to
the right sector is $k'_s=k'_b+2=2$, and we have $c^{\rm
  ws}_R(S^3/\ZZ_{N}) = \frac{3}2$. We are thus left with a trivial
theory in the left sector and three fermions $\bar \chi^a$ ($a=1,2,3$)
in the right sector. The AdS$_3$ part of the geometry is described by
a heterotic $SL(2)$ WZW model with levels $k_b=4$ and $k_s=2$. The
full (supersymmetric part of the) background is thus
\begin{align}
SL(2, \RR)_2 \times \{\bar \chi^1, \bar\chi^2,\bar\chi^3 \} \times T^4 \,,
\end{align} 
and the central charges of the worldsheet model are:
\begin{align}
c^{\rm ws}_L(SL(2))=6 \,,&\qquad c^{\rm ws}_L(S^3/\ZZ_{N})=0\,,\qquad
c^{\rm ws}_L(T^4)=4 \,,\nn\\
c^{\rm ws}_R(SL(2))=15/2 \,,&\qquad c^{\rm
  ws}_R(S^3/\ZZ_{N})=3/2\,,\qquad c^{\rm ws}_R(T^4)=6 \,,
\end{align}
ensuring criticality, $(c^{\rm ws}_L, c^{\rm ws}_R)=(26,15)$, given
that $c^{\rm ws}_L(E_8 \times
E_8)=16$.  The worldsheet model also gives the correct central charges
for the boundary CFT, cf.\ Eq.~(\ref{cc3}). Related heterotic models
involving three fermions can be found in \cite{Giveon:2006pr,
  Dabholkar}.

We conclude with some comments on the dual boundary conformal field
theory.  Removing the D5 branes in the quiver ADHM theory corresponds
to the removal of the outer circle and the spikes in the quiver
diagram in figure~\ref{quiverfig}. The ADHM part of the quiver action
disappears, leaving only that part of the action which corresponds to
the inner circle of the quiver diagram. Nevertheless, the counting of
the massless degrees of freedom in the remaining quiver theory seems
to yield the correct central charges, $(c_L, c_R)=(24p, 12 p)$ (for $N
\geq 2$).  It is interesting to observe that the independence of
$c_{L,R}$ on $N$ is reflected by fact that varying $N$ changes only
the number of sites in the quiver diagram corresponding to $U(p)$
gauge groups. Recall, however, that the fields of the $U(p)$ gauge
groups do not contribute to the central charges of the infrared
conformal field theory. Certainly, it would be interesting to study
this field theory in more detail.

\subsection{Heterotic NS5-KKM intersection (\texorpdfstring{$p=0$}{})}

For completeness, we also consider the NS5-KKM intersection which can
be obtained from the three-charge model of section~\ref{sec21} by
setting $p=0$.

Let us approach this setup from a slightly different point of view.
In \cite{Strominger} Lapan, Simons and Strominger suggested to start
from a four-dimensional monopole black hole with near-horizon geometry
\begin{align}
\RR^t \times \RR^\phi \times S^2 \times T^6\,,
\end{align}
where $\RR^t$ denotes time and $\RR^\phi$ a real line labelled by
$\phi$ with linear dilaton.  Decompactifying one of the compact
directions, \ie replacing $\RR^t \times S^1$ by a two-dimensional
Minkowski space $\RR^{1,1}$ leads to the geometry
\begin{align}
\RR^{1,1} \times \RR^\phi \times S^2 \times T^5 \,. \label{geomMS}
\end{align}
The CFT on (\ref{geomMS}) is then expected to describe a monopole
string in five dimensions \cite{Strominger}. Ref.~\cite{Strominger}
also suggested that the $S^2$ factor could be described by the
coset model of
\cite{Giddings}.

Here, however, we deviate from the proposal of \cite{Strominger} and
include a KK monopole charge by replacing $S^2 \times T^5$ by
$S^3/\ZZ_{N} \times T^4$. Of course, we thereby break half of the
target space supersymmetry.  Heterotic string theory in the background of
a five-dimensional monopole string with additional KK monopole charge
is then expected to be given by the CFT on
\begin{align}
\RR^{1,1} \times \RR^\phi \times S^3/\ZZ_{N} \times T^4 \,. 
\label{geom}
\end{align}
In fact, the thus derived background is nothing but the
near-horizon geometry of the F1-NS5-KKM set-up for {\em vanishing}
electrical F1 charge, $p=0$. This can be seen by setting $F=1$ in
(\ref{classicalmetric}) and taking the limit $r \rightarrow 0$.

Heterotic string theory on the background (\ref{geom}) can
be described by a linear dilaton theory with central charges
\begin{align} \label{cdilaton}
c^{\rm ws}_L(\RR^{1,1} \times \RR^\phi)=2+(1+3 Q_D^2)\,,\qquad
c^{\rm ws}_R(\RR^{1,1} \times \RR^\phi)=3+(\textstyle\frac{3}{2}+3 Q_D^2) \,,
\end{align}
and dilaton charge $Q_D$. The internal part of the geometry,
$S^3/\ZZ_{N}$ and $T^4$, will be described as before, see section~2.3.
By criticality, the linear dilaton charge $Q_D$ is related to the
bosonic level $k_b'$ of the $S^3/\ZZ_{N}$ theory as
\begin{align}
Q_D^2 = \frac{2}{k_b' +2} \,,
\end{align}
where $k'_b=k'_s-2 = N N'$, if we assume $k'_s=NN'+2$. 

\medskip
Finally, as explained in \cite{Giveon:2005mi}, there is a simple
relation between linear dilaton and $SL(2)$ models. Adding $p$
D1-branes along the $R^{1,1}$ and taking the near-horizon limit
amounts to replacing the factor $\RR^{1,1} \times \RR^\phi$ by
$AdS_3$. The level of $SL(2)$ is related to the dilaton charge 
by $k_s=2/Q_D^2$ ($k_b=k_s+2$).  This leads back to $AdS_3 \times
S^3/\ZZ_{N} \times T^4$, as expected.

\setcounter{equation}{0}
\section{Conclusions}

We studied the AdS$_3$/CFT$_2$ correspondence of a heterotic
three-charge model with ($0, 4$) supersymmetry. We gathered evidence
for the equivalence of the following two theories:
\begin{itemize}
\item[i)] $E_8 \times E_8$ heterotic string theory on $AdS_3 \times
  S^3/\ZZ_N \times T^4$ 
\item[ii)] the $(0, 4)$ Higgs branch theory of a $\ZZ_N$ orbifold of
  Witten's ADHM sigma model
\end{itemize} 
We motivated the duality by studying the low-energy effective action
of a particular type~I setup dual to a heterotic configuration with
$AdS_3 \times S^3/\ZZ_N \times T^4$ near-horizon geometry.  We
constructed the ultraviolet theory in terms of a $\ZZ_N$ orbifold of
the ADHM massive sigma model \cite{Witten:1994tz} and verified that
the corresponding Higgs branch theory has the correct central charges.
We also found that the first-order $\alpha'$-corrected supergravity
solution correctly reproduces the (supersymmetric) central charge of
the boundary conformal field theory up to terms of order ${\cal
O}(\frac{1}{NN'})$, cf.~(\ref{cccorr}) with (\ref{cc1}).

\medskip The proposed heterotic duality obviously requires further
investigation. The evidence we gave is based on the counting of the
massless degrees of freedom of the ultraviolet orbifold theory.
These modes are not renormalised and therefore also constitute the
Higgs branch theory. Its actual construction is expected to be
straightforward along the lines of \cite{Lambert} by integrating out
the massive modes in the UV theory. This procedure will be made more 
complicated by
the fact that the Higgs branch metric will receive $\alpha'$
corrections and seems to be divergent at the origin \cite{Lambert}. It
would also be interesting to work out the dictionary between the
chiral primaries of the boundary CFT and those of the worldsheet
model~\cite{KLL}. The primaries of the boundary CFT will be composite
operators of the massless fields of the ultraviolet ADHM quiver
model. A comparison of the corresponding $n$-point functions should
then provide further evidence for the duality. Such tests have
previously been performed in the type~II AdS$_3$/CFT$_2$ duality in
\cite{Gaberdiel, Pakman1, Pakman2, Taylor:2007hs, Giribet:2007wp}.

\subsection*{Acknowledgements}

We would like to thank Ilka Brunner, David Kutasov and Finn Larsen for
helpful discussions and comments related to this work. Moreover, we
are deeply indebted to Matthias Gaberdiel, Amit Giveon and Neil
Lambert for invaluable comments on a preliminary version of this
paper.  I.K.\ is grateful to Angelo Lopez for an extended exchange of
emails on ample divisors of Calabi-Yau manifolds.
This research has been partially supported by the Swiss National Science
Foundation and the Marie Curie network `Constituents, Fundamental Forces and
Symmetries of the Universe' (MRTN-CT-2004-005104).


\newpage
\appendix
\noindent {\LARGE \bf Appendix}

\renewcommand{\theequation}{\thesection.\arabic{equation}}
\setcounter{equation}{0}
\section{Web of Dualities}\label{dualitiesMtheory}
In this appendix we display the various dualities leading from the
heterotic theory on $T^5$ (along $x^{5,6,7,8,9}$) to M-theory on
$K3\times T^2$ (along $x^{6,7,8,9}$ and $x^{5,10}$). For a review of
string dualities see e.g.~\cite{Antoniadis:1999yx}. In order to
facilitate keeping track of the various steps, we have depicted a
schematic overview in the following web
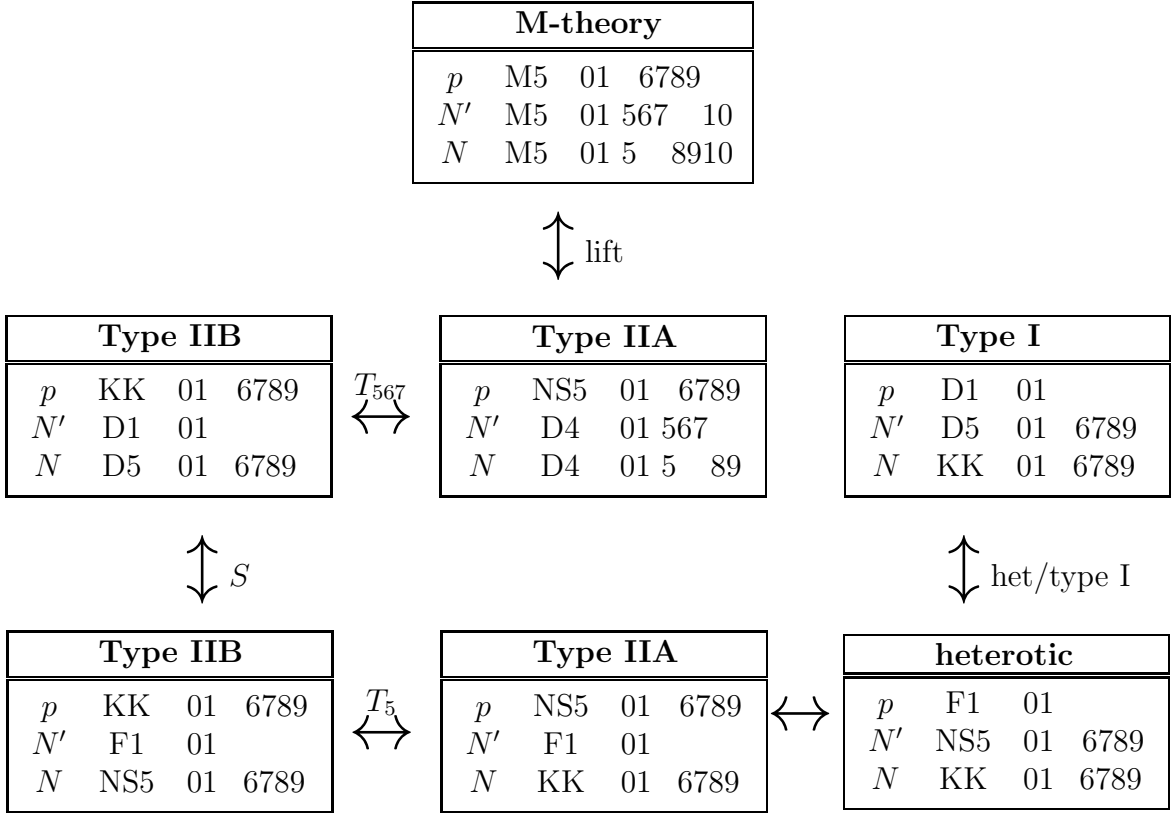
\begin{center}
\fbox{\parbox{4.3cm}{\hspace{1.25cm}\textbf{M-theory}}}\\
\fbox{\parbox{4.3cm}{
\begin{tabular}{ccl}
$p$ & M5 &01\hspace{0.37cm}6789\\
$N'$ & M5 &01 567\hspace{0.45cm}10\\
$N$ & M5 &01 5\hspace{0.45cm}8910
\end{tabular}}}\\[10pt]
{\huge $\updownarrow$} {lift}
\end{center}
\begin{center}
\parbox{4.2cm}{\fbox{\parbox{4.1cm}{\hspace{1.1cm}\textbf{Type IIB}}}\\
\fbox{\parbox{4.1cm}{
\begin{tabular}{ccl}
$p$ & KK &01\hspace{0.37cm}6789\\
$N'$ & D1 &01 \\
$N$ & D5 &01 \hspace{0.2cm}6789
\end{tabular}}}}
\begin{tabular}{c} $T_{567}$ \\ {\huge $\leftrightarrow$} \end{tabular}
\parbox{4.2cm}{\fbox{\parbox{4.1cm}{\hspace{1.1cm}\textbf{Type IIA}}}\\
\fbox{\parbox{4.1cm}{
\begin{tabular}{ccl}
$p$ & NS5 &01\hspace{0.37cm}6789\\
$N'$ & D4 &01 567\\
$N$ & D4 &01 5\hspace{0.45cm}89
\end{tabular}}}}
{\hspace{0.9cm}}
\parbox{4.2cm}{\fbox{\parbox{4.1cm}{\hspace{1.1cm}\textbf{Type I}}}\\
\fbox{\parbox{4.1cm}{
\begin{tabular}{ccl}
$p$ & D1 &01\\
$N'$ & D5 &01\hspace{0.37cm}6789\\
$N$ & KK &01 \hspace{0.2cm}6789\\
\end{tabular}}}}
{\hspace{1.2cm}}
\parbox{4.2cm}{\hspace{4.1cm}}\\[10pt]
\hspace{2cm}{\huge{$\updownarrow$}} $S$ \hspace{9cm}
{\huge $\updownarrow$} {het/type I}
\begin{center}
\parbox{4.2cm}{\fbox{\parbox{4.1cm}{\hspace{1.1cm}\textbf{Type IIB}}}\\
\fbox{\parbox{4.1cm}{
\begin{tabular}{ccl}
$p$ & KK &01\hspace{0.37cm}6789\\
$N'$ & F1 &01 \\
$N$ & NS5 &01 \hspace{0.2cm}6789
\end{tabular}}}}
\begin{tabular}{c} $T_5$ \\ {\huge $\leftrightarrow$} \end{tabular}
\parbox{4.2cm}{\fbox{\parbox{4.1cm}{\hspace{1.1cm}\textbf{Type IIA}}}\\
\fbox{\parbox{4.1cm}{
\begin{tabular}{ccl}
$p$ & NS5 &01\hspace{0.37cm}6789\\
$N'$ & F1 &01 \\
$N$ & KK &01 \hspace{0.2cm}6789
\end{tabular}}}}
{\huge$\leftrightarrow$} 
\parbox{4.2cm}{\fbox{\parbox{4.1cm}{\hspace{1.1cm}\textbf{heterotic}}}\\
\fbox{\parbox{4.1cm}{
\begin{tabular}{ccl}
$p$ & F1 &01\\
$N'$ & NS5 &01\hspace{0.37cm}6789\\
$N$ & KK &01 \hspace{0.2cm}6789\\
\end{tabular}}}}
\end{center}
\end{center}
We start in the lower right corner with the heterotic theory as
described in section \ref{sec21}. Following the first arrow to the
left\footnote{The arrow pointing upwards is just included for
  completeness and represents the heterotic-type~I duality which we
  exploit in section \ref{bcftmodel}.}, heterotic-type IIA duality
takes us to a setup with NS5-branes, fundamental strings and KK
monopoles as described in the corresponding box. Going further to the
left (using the arrow labelled $T_5$), we perform a T-duality along
the isometry direction of the KK monopoles (direction $x^5$), which
exchanges the KK monopoles and the NS5-branes but leaves the F1
untouched. Since we have performed the T-duality only along a single
direction, the setup is now in the type IIB theory. Following the next
arrow upwards (labelled by S), we perform S-duality in the type IIB
framework, which turns the NS5-branes and F1 into D5- and D1-branes,
respectively. Next we follow the arrow labelled $T_{567}$ to the
right, which represents T-duality transformations along $x^{5,6,7}$.
Since again the isometry direction of the KK monopoles is affected,
they are transformed to NS5-branes, while the D1 and D5-branes are
mapped to D4-branes. Since we have performed the duality
transformation in an odd number of dimensions, we are back to the type
IIA framework. The final arrow pointing upwards is the M-theory lift,
which takes us to the setup of three stacks of M5-branes described in
section \ref{sec22}.



\begin{thebibliography}{99}

\bibitem{Giveon:2006pr} A.~Giveon and D.~Kutasov, {\it Fundamental strings and black holes,} JHEP {\bf 0701}, 071 (2007) [arXiv:hep-th/0611062].
\bibitem{TalkStrominger} A.~Strominger, {\it Search for the
    Holographic Dual of N Heterotic Strings}, Talk given at
  Strings~2007, Madrid, Spain, June 2007.

\bibitem{Dabholkar} A.~Dabholkar and S.~Murthy, {\it Fundamental Superstrings as Holograms,}  JHEP {\bf 0802}, 034 (2008) [arXiv:0707.3818 [hep-th]].
\bibitem{Johnson2} C.~V.~Johnson, {\it Heterotic Coset Models of Microscopic Strings and Black Holes,} arXiv:0707.4303 [hep-th].
\bibitem{Strominger} J.~M.~Lapan, A.~Simons and A.~Strominger, {\it Nearing the Horizon of a Heterotic String,} arXiv:0708.0016 [hep-th]
\bibitem{Kraus} P.~Kraus, F.~Larsen and A.~Shah, {\it Fundamental Strings, Holography, and Nonlinear Superconformal Algebras,}  JHEP {\bf 0711}, 028 (2007)
[arXiv:0708.1001 [hep-th]].
\bibitem{Alishahiha2008}
  M.~Alishahiha and S.~Mukhopadhyay,
  {\it On Six Dimensional Fundamental Superstrings as Holograms,}
  arXiv:0803.0685 [hep-th].

\bibitem{CDKL} A.~Castro, J.~L.~Davis, P.~Kraus and F.~Larsen, {\it 5D attractors with higher derivatives,} JHEP {\bf 0704}, 091 (2007) [arXiv:hep-th/0702072]; A.~Castro, J.~L.~Davis, P.~Kraus and F.~Larsen, {\it 5D Black Holes and Strings with Higher Derivatives,} JHEP {\bf 0706}, 007 (2007) [arXiv:hep-th/0703087].

\bibitem{Castro:2008ne} A.~Castro, J.~L.~Davis, P.~Kraus and F.~Larsen, {\it String Theory Effects on Five-Dimensional Black Hole Physics,} arXiv:0801.1863 [hep-th].

\bibitem{Alishahiha2007}
  M.~Alishahiha, F.~Ardalan, H.~Ebrahim and S.~Mukhopadhyay,
  {\it On 5D Small Black Holes,}
  JHEP {\bf 0803}, 074 (2008) [arXiv:0712.4070 [hep-th]].

\bibitem{KLL} D.~Kutasov, F.~Larsen and R.~G.~Leigh, {\it String theory in magnetic monopole backgrounds,} Nucl.\ Phys.\  B {\bf 550}, 183 (1999) [arXiv:hep-th/9812027].

\bibitem{Witten:1994tz} E.~Witten, {\it Sigma Models And The ADHM Construction Of Instantons,} J.\ Geom.\ Phys.\  {\bf 15}, 215 (1995) [arXiv:hep-th/9410052].
\bibitem{Douglas} M.~R.~Douglas, {\it Gauge Fields and D-branes,} J.\ Geom.\ Phys.\  {\bf 28}, 255 (1998) [arXiv: hep-th/9604198].
\bibitem{Sugawara} Y.~Sugawara, {\it N = (0,4) quiver SCFT(2) and supergravity on AdS(3) x S(2),} JHEP {\bf 9906}, 035 (1999) [arXiv:hep-th/9903120].
\bibitem{Okuyama:2005gq}
  K.~Okuyama, {\it D1-D5 on ALE space,}  JHEP {\bf 0512} (2005) 042
  [arXiv:hep-th/0510195].

\bibitem{Lambert2} N.~D.~Lambert, {\it Quantizing the (0,4) supersymmetric ADHM sigma model,} Nucl.\ Phys.\  B {\bf 460}, 221 (1996) [arXiv:hep-th/9508039].
\bibitem{Lambert} N.~D.~Lambert, {\it D-brane bound states and the generalised ADHM construction,} Nucl.\ Phys.\  B {\bf 519}, 214 (1998) [arXiv:hep-th/9707156].

\bibitem{Metsaev:1987zx} R.~R.~Metsaev and A.~A.~Tseytlin, {\it Order alpha-prime (Two Loop) Equivalence of the String Equations of Motion and the Sigma Model Weyl Invariance Conditions: Dependence on the Dilaton and the Antisymmetric Tensor,} Nucl.\ Phys.\  B {\bf 293} (1987) 385.

\bibitem{Hull:1987pc} C.~M.~Hull and P.~K.~Townsend, {\it The Two Loop Beta Function For Sigma Models With Torsion,} Phys.\ Lett.\  B {\bf 191} (1987) 115.

\bibitem{Sen:2005wa} A.~Sen, {\it Black hole entropy function and the attractor mechanism in higher derivative gravity,} JHEP {\bf 0509} (2005) 038 [arXiv:hep-th/0506177].
\bibitem{Sen:2005iz} A.~Sen, {\it Entropy function for heterotic black holes,} JHEP {\bf 0603} (2006) 008 [arXiv:hep-th/0508042].

\bibitem{BH} J.~D.~Brown and M.~Henneaux, {\it Central Charges in the Canonical Realization of Asymptotic Symmetries: An Example from Three-Dimensional Gravity,} Commun.\ Math.\ Phys.\  {\bf 104}, 207 (1986).

\bibitem{Maldacena} J.~M.~Maldacena, A.~Strominger and E.~Witten, {\it Black hole entropy in M-theory,} JHEP {\bf 9712}, 002 (1997) [arXiv:hep-th/9711053].
\bibitem{Giddings} S.~B.~Giddings, J.~Polchinski and A.~Strominger, {\it Four-dimensional black holes in string theory,} Phys.\ Rev.\  D {\bf 48}, 5784 (1993) [arXiv:hep-th/9305083].
\bibitem{Seiberg} A.~Giveon, D.~Kutasov and N.~Seiberg, {\it Comments on string theory on AdS(3),} Adv.\ Theor.\ Math.\ Phys.\  {\bf 2}, 733 (1998) [arXiv:hep-th/9806194].
\bibitem{Lowe} D.~A.~Lowe, {\it E(8) x E(8) small instantons in matrix theory,} Nucl.\ Phys.\  B {\bf 519}, 180 (1998) [arXiv:hep-th/9709015].

\bibitem{Douglas:1995bn} M.~R.~Douglas, {\it Branes within branes,} arXiv:hep-th/9512077.
\bibitem{Witten10} E.~Witten, {\it Small Instantons in String Theory,} Nucl.\ Phys.\  B {\bf 460}, 541 (1996) [arXiv:hep-th/9511030].
\bibitem{Douglas:1996sw} M.~R.~Douglas and G.~W.~Moore, {\it D-branes, Quivers, and ALE Instantons,} arXiv:hep-th/9603167.
\bibitem{Johnson:1996py}
  C.~V.~Johnson and R.~C.~Myers, {\it Aspects of type IIB theory on ALE spaces,} Phys.\ Rev.\  D {\bf 55}, 6382 (1997) [arXiv:hep-th/9610140].
\bibitem{Johnson} C.~V.~Johnson, {\it Anatomy of a duality,} Nucl.\ Phys.\  B {\bf 521}, 71 (1998) [arXiv:hep-th/9711082].

\bibitem{Gimon:1996rq}
  E.~G.~Gimon and J.~Polchinski, {\it Consistency Conditions for Orientifolds and D-Manifolds,} Phys.\ Rev.\  D {\bf 54}, 1667 (1996) [arXiv:hep-th/9601038].

\bibitem{Constable:2002vt} N.~R.~Constable, J.~Erdmenger, Z.~Guralnik and I.~Kirsch, {\it (De)constructing intersecting M5-branes,} Phys.\ Rev.\  D {\bf 67}, 106005 (2003) [arXiv:hep-th/0212136].
\bibitem{Barbon:1998nx} J.~L.~F.~Barbon, J.~L.~Manes and M.~A.~Vazquez-Mozo, {\it Large N limit of extremal non-supersymmetric black holes,} Nucl.\ Phys.\  B {\bf 536}, 279 (1998) [arXiv:hep-th/9805154].

\bibitem{Hanaki:2006pj} K.~Hanaki, K.~Ohashi and Y.~Tachikawa, {\it Supersymmetric Completion of an $R^2$ Term in Five-Dimensional Supergravity,} Prog.\ Theor.\ Phys.\  {\bf 117} (2007) 533 [arXiv:hep-th/0611329].

\bibitem{Sen:2007qy} A.~Sen, {\it Black Hole Entropy Function, Attractors and Precision Counting of Microstates,} arXiv:0708.1270 [hep-th].

\bibitem{Garousi:2007zb} M.~R.~Garousi and A.~Ghodsi, {\it On Attractor Mechanism and Entropy Function for Non-extremal Black Holes/Branes,} JHEP {\bf 0705} (2007) 043 [arXiv:hep-th/0703260]. M.~R.~Garousi and A.~Ghodsi, {\it Entropy Function for Non-extremal D1D5 and D2D6NS5-branes,} JHEP {\bf 0710} (2007) 036 [arXiv:0705.2149 [hep-th]].

\bibitem{Sahoo:2006pm} B.~Sahoo and A.~Sen, {\it alpha'-Corrections to
    Extremal Dyonic Black Holes in Heterotic String Theory,} JHEP {\bf
    0701} (2007) 010 [arXiv:hep-th/0608182]. 
\bibitem{Antoniadis:1999yx} I.~Antoniadis and G.~Ovarlez, {\it An introduction to perturbative and non-perturbative string theory,} arXiv:hep-th/9906108.

\bibitem{Giveon:2005mi} A.~Giveon, D.~Kutasov, E.~Rabinovici and A.~Sever, {\it Phases of quantum gravity in AdS(3) and linear dilaton backgrounds,} Nucl.\ Phys.\  B {\bf 719}, 3 (2005) [arXiv:hep-th/0503121].

\bibitem{Gaberdiel} M.~R.~Gaberdiel and I.~Kirsch, {\it Worldsheet correlators in AdS(3)/CFT(2),} JHEP {\bf 0704}, 050 (2007) [arXiv:hep-th/0703001].
\bibitem{Pakman1} A.~Dabholkar and A.~Pakman, {\it Exact chiral ring of AdS(3)/CFT(2),} arXiv:hep-th/0703022.
\bibitem{Pakman2} A.~Pakman and A.~Sever, {\it Exact N=4 correlators of AdS(3)/CFT(2),} Phys.\ Lett.\  B {\bf 652}, 60 (2007) [arXiv:0704.3040 [hep-th]].
\bibitem{Taylor:2007hs} M.~Taylor, {\it Matching of correlators in $AdS_3/CFT_2$,} JHEP {\bf 0806}, 010 (2008)
  [arXiv:0709.1838 [hep-th]].

\bibitem{Giribet:2007wp} G.~Giribet, A.~Pakman and L.~Rastelli, {\it
    Spectral Flow in AdS(3)/CFT(2),} JHEP {\bf 0806}, 013 (2008)
  [arXiv:0712.3046 [hep-th]].



\end{thebibliography}
\end{document}